\documentclass[preprint,12pt]{elsarticle}

\usepackage{microtype}
\usepackage{amssymb}
\usepackage[margin=2.5cm]{geometry} %1 inch margin
\usepackage{lipsum}
\usepackage{graphicx}
\usepackage{caption}
\usepackage{subcaption}
\usepackage{graphicx}
\usepackage{booktabs, tabularx, enumitem, threeparttable}
\usepackage{booktabs}

\usepackage{multirow}
\usepackage{float} 
\usepackage{makecell}
\usepackage{subcaption}
\usepackage{threeparttable}
\usepackage[symbol]{footmisc}
\usepackage[usenames,dvipsnames]{color}
\usepackage[hidelinks]{hyperref}
\usepackage{amssymb}
\usepackage{geometry}
\usepackage{rotating}
\usepackage{graphicx}
\usepackage{adjustbox}
\usepackage{pdflscape}
\usepackage[utf8]{inputenc}
\usepackage{array}
\usepackage{booktabs}
\usepackage{graphicx}
\usepackage{float}
\usepackage{makecell}

\usepackage{diagbox}
\usepackage{multirow}
\usepackage{booktabs}
\usepackage{diagbox}
\usepackage{amsmath}
\usepackage{url}
\usepackage{hyperref}
\usepackage{enumitem}
\usepackage{xurl}
\usepackage{multirow} 

\usepackage[usenames,dvipsnames]{color}
\usepackage{multirow}
\usepackage{amssymb}
\usepackage{amsmath}
\usepackage{graphicx}
\usepackage{float} % for H specifier
\usepackage{lineno}
\usepackage{natbib}

\setcitestyle{square}

\makeatletter
 \def\ps@pprintTitle{%
  \let\@oddhead\@empty
  \let\@evenhead\@empty
  \def\@oddfoot{}%
  \let\@evenfoot\@oddfoot}
 \makeatother
\begin{document}
\begin{frontmatter}

%% Title, authors and addresses

%% use the tnoteref command within \title for footnotes;
%% use the tnotetext command for theassociated footnote;
%% use the fnref command within \author or \affiliation for footnotes;
%% use the fntext command for theassociated footnote;
%% use the corref command within \author for corresponding author footnotes;
%% use the cortext command for theassociated footnote;
%% use the ead command for the email address,
%% and the form \ead[url] for the home page:
%% \title{Title\tnoteref{label1}}
%% \tnotetext[label1]{}
%% \author{Name\corref{cor1}\fnref{label2}}
%% \ead{email address}
%% \ead[url]{home page}
%% \fntext[label2]{}
%% \cortext[cor1]{}
%% \affiliation{organization={},
%%            addressline={}, 
%%            city={},
%%            postcode={}, 
%%            state={},
%%            country={}}
%% \fntext[label3]{}

\title{Persona Migration and Expectation Recalibration in Generative AI Adoption: A Longitudinal Study at a State Department of Transportation}

\author[inst1]{Omidreza Shoghli}
\author[inst1]{Fatemeh Banani Ardecani}
\author [inst2] {Amin Mohamadi Hezaveh}
% Author affiliation
\affiliation[inst1]{organization={William States Lee College of Engineering, University of North Carolina at Charlotte},%Department and Organization
            addressline={9201 University City Blvd}, 
            city={Charlotte},
            postcode={28223}, 
            state={North Carolina},
            country={USA}}
\affiliation[inst2]{organization={North Carolina Department of Transportation},%Department and Organization
            city={Raleigh},
            postcode={27699}, 
            state={North Carolina},
            country={USA}}

%% Abstract
\begin{abstract} 
Generative AI tools are increasingly being piloted in public agencies, yet limited evidence explains how employee acceptance evolves after hands-on use. This study examines Microsoft 365 Copilot adoption during an eight-week pilot at a state Department of Transportation. Using a matched two-wave survey, we measured perceived usefulness, perceived ease of use, behavioral intention, and trust before training and access and again after pilot participation. After matching and response quality screening, the analytic sample included 124 employees. Nonparametric pre-post tests assessed aggregate construct changes, while k-means clustering identified baseline acceptance personas and fixed-centroid assignment tracked post-pilot migration. Open-ended responses were analyzed using exploratory keyword-based content mapping.
At the aggregate level, perceived usefulness declined significantly after hands-on use, suggesting recalibration of initial expectations, a pattern consistent with negative disconfirmation as described in Bhattacherjee's Expectation–Confirmation Model of IT continuance. Perceived ease of use, behavioral intention, and trust showed only small, non-significant changes. Cluster analysis identified three baseline personas: Skeptics, Cautiously Positive users, and Champions. Although overall persona counts changed only modestly, individual movement was substantial. Forty percent of baseline Skeptics moved to Cautiously Positive, while 68\% of baseline Champions moved to less enthusiastic personas. Migration paths showed that upward movement was associated with gains in usefulness, behavioral intention, and trust, whereas downward movement was associated with declines in usefulness and trust. Task-use patterns further indicated that communication and summarization remained comparatively stable use cases, while data/chart and presentation tasks declined after experience. Concern patterns shifted as accuracy and privacy/data concerns decreased, but job/skills concerns increased.
Findings suggest that public sector AI adoption should be monitored dynamically and supported through persona-specific training, workflow-specific examples, verification routines, and trust calibration safeguards. The study contributes a practical framework for tracking workforce heterogeneity during enterprise generative AI implementation efforts.
\end{abstract}

%%Research highlights
%\begin{highlights}

%\item Tracked Microsoft 365 Copilot acceptance over an 8-week DOT pilot (n=124)
%\item Perceived usefulness declined significantly after hands-on use
%\item K-means identified three baseline acceptance personas with substantial migration
%\item 68\% of Champions migrated to less enthusiastic personas after pilot exposure
%\item Trust calibration emerged as central to sustained generative AI adoption

%\end{highlights}


\end{frontmatter}

%% Add \usepackage{lineno} before \begin{document} and uncomment 
%% following line to enable line numbers
%% \linenumbers

%% main text

%% Use \section commands to start a section 

\section{Introduction and Background}
Generative AI has rapidly emerged as one of the most disruptive technologies, acting as a catalyst for digital transformation across various industries \cite{Bies2024FutureOffice}. 
Recent economic analyses estimate that up to 40\% of occupations could be impacted by generative AI within the next few years \cite{MorganStanley_TMT_2023, Eloundou2023GPTs, orchard2023rise}. This potential spans knowledge-intensive tasks such as drafting documents, analyzing data, and customer service interactions. Public sector organizations are increasingly recognizing generative AI's promise to enhance productivity and decision-making \cite{BCG2024PublicSector}. In particular, state Departments of Transportation (DOTs) are exploring tools such as Microsoft 365 Copilot, an AI assistant that integrates large language models (LLMs) into everyday office applications. By offloading routine but time-consuming tasks, DOTs hope to redirect staff effort toward higher-value engineering, planning, and stakeholder coordination.

\subsection{Enterprise Adoption of Generative AI}
In 2023, less than a year after ChatGPT's launch, it was reported \cite{McKinsey2023StateAI} that one-third of companies regularly use generative AI in at least one function, and nearly one-quarter of C-suite executives are personally using these tools for work. Momentum is backed by budget with 92\% of companies planning to raise AI spending over the next three years \cite{McKinsey2025Superagency}. Early pilot studies indicate that LLM tools raise self‑reported productivity and work quality among the majority of users \cite{microsoft_work_trend_index_2024, Microsoft2023Copilot, Bies2024FutureOffice}.
Typical quick‑win use cases include drafting technical memos, summarizing project meetings, and generating slide decks.
Public agencies are moving in parallel. A 2024 NASCIO survey shows that US states have 
initiated generative AI efforts, with the majority having completed research, issued risk guardrails, collaborated with the private sector and invested in training \cite{NASCIO_McKinsey_2024_GenAI}. State DOTs are also testing generative AI tools to relieve personnel of low‑value tasks and to accelerate project communication, an organizational shift that ultimately depends on how employees perceive the tool's usefulness, ease of use, trustworthiness, and overall value \cite{Heo2025AEC}.
Despite optimism, enterprises face governance and implementation challenges as they adopt LLM tools. Current trends reveal a gap between swift experimentation and the slower development of risk management practices \cite{Beltran2024GovRisk}. Fewer than half of organizations using AI have measures in place to mitigate risk of generative AI, such as output inaccuracy (e.g. hallucinated facts) \cite{McKinsey2023StateAI}, indicating that many organizations are still learning how to balance innovation with oversight.

Another challenge accompanying enterprise AI adoption is workforce readiness, including user trust, skills, and perceptions of the technology's usefulness and ease of use. For many organizations shortages of staff who know how to use or manage AI are the top barrier to wider deployment \cite{NASCIO_McKinsey_2024_GenAI} and most firms expect to spend on retraining so employees can write effective prompts, verify AI outputs, and shift time to higher‑value tasks \cite{McKinsey2025Superagency}. 
In the context of AI systems, trust is an important factor in adoption and user behavior \cite{Wester2024LLMAdvice, Kumar2025AIScooters}. Insufficient trust can suppress utilization, while uncritical trust may permit inaccurate or biased outputs \cite{Choudhury2023TrustChatGPT}. These concerns are amplified in public sector settings, where information produced by AI must meet stringent accuracy and fairness standards.

\subsection{Research Gaps in Public-Sector AI Adoption}
Although pilot projects and early surveys provide a first look at AI uptake, several critical questions remain unanswered, particularly for mission‑driven agencies such as state DOTs. (1) Public sector workforces are heterogeneous, comprising employees of diverse roles (technical, administrative, field, office, etc.) and varying levels of technical proficiency. This heterogeneity means a one-size-fits-all analysis of the average user's response to the adoption of LLM tools may obscure important subgroup differences. Segmenting employees by the four constructs drawn from Technology Acceptance Model \cite{Davis1989TAM}, perceived usefulness (PU), perceived ease of use (PEOU), behavioral intention (BI), and trust (TR), can reveal distinct user profiles whose training and support requirements are different. Current public‑sector literature provides little guidance on how to identify or monitor these segments. (2) Most studies capture user perceptions at a single time-point. There is a lack of longitudinal research tracking how initial excitement or concerns change after real-world use or training. This is particularly relevant for LLM applications, where user perceptions of usefulness, ease, and trust may shift as they gain hands-on experience or encounter the tool's limitations. Monitoring these changes over time is crucial to formulating effective implementation strategies such as timely refresher trainings. As noted in recent literature, user acceptance of AI is not a static phenomenon; it involves an ongoing learning curve and adaptation process, which researchers are only beginning to document \cite{10.3389/frai.2025.1565927, Wester2024LLMAdvice}. 
A related theoretical gap concerns the frameworks used to interpret post-use perceptions. The Technology Acceptance Model was originally validated in pre-use and early-use contexts, where perceived usefulness and perceived ease of use reflect anticipated performance rather than confirmed experience \cite{Davis1989TAM}. For post-use measurement, Bhattacherjee's Expectation–Confirmation Model of IT continuance (ECM-IT) offers a complementary lens: users form pre-use expectations, evaluate actual performance through use, and experience satisfaction or disconfirmation that in turn shapes continuance intention \cite{bhattacherjee2001understanding}. Where TAM predicts adoption of a new system, ECM-IT predicts continued use following initial exposure, precisely the transition this study examines. Applying both frameworks in tandem allows us to test not only whether employees intend to adopt Copilot, but whether their post-pilot perceptions reflect confirmation or disconfirmation of their baseline expectations.
Finally, there is a gap in translating technology acceptance insights into practical adoption strategies for public agencies. Questions remain open on how to effectively guide a public-sector workforce through the adoption curve of generative AI and how can trust in AI outputs be enhanced or mistrust allayed through policy or interface design? 
This study aims to fill these gaps by closely monitoring a state DOT's experience with an LLM tool over time, from pre-deployment to post-training and use.

\subsection{Study Contributions and Approach}
This paper addresses the above gaps by presenting a longitudinal, user-centered investigation of large language model tool adoption in a state DOT. 
We draw on a matched pre- and post-pilot survey of NCDOT employees introduced to Microsoft~365 Copilot as part of an eight-week AI pilot. Of the pilot participants, 175 completed the pre-pilot survey and 133 completed both waves. After data quality screening, the final analytic sample comprised 124 matched respondents. Using the TAM framework augmented with a trust construct, we measure key acceptance variables, Perceived Usefulness (PU), Perceived Ease of Use (PEOU), Behavioral Intention (BI), and Trust (TR), before the Copilot training and again several weeks after employees went through training and began using the tool. This design enables us to quantify shifts in employee perceptions and intentions over time and to test which changes are significant. To our knowledge, this is one of the first longitudinal studies of generative AI acceptance within a public transportation agency, offering insight into how initial attitudes evolve with real-world exposure.
In addition, we employ cluster analysis to uncover latent user segments based on the TAM and trust variables. Rather than treating the workforce as homogeneous, we identify distinct clusters (user profiles). For example, a cluster of Enthusiasts who report high usefulness, ease, and trust, versus a cluster of Skeptics who report low usefulness and trust. We then analyze cluster migration between the pre- and post-pilot surveys, tracking how individuals moved from one profile to another. This novel approach reveals patterns of adoption readiness and resistance: for instance, we can see what fraction of initially skeptical users became more positive after hands-on experience, and conversely whether any initially enthusiastic users grew cautious. Understanding these transitions is valuable for pinpointing which groups may need additional support or see diminishing perceived value over time. The clustering also allows us to map workforce heterogeneity in adoption – capturing the fact that even at a single point in time, employees are at different stages of the acceptance curve.

Overall, our study makes the following contributions: 
(1) We provide empirical data on how TAM constructs (PU, PEOU, BI) and trust in an AI assistant change over time with training and usage, and we interpret the observed post-use shifts through the complementary lens of Bhattacherjee's Expectation–Confirmation Model of IT continuance. We use ECM-IT as an interpretive lens for reading the directional post-pilot shifts in PU, BI, and trust as positive or negative disconfirmation. This combination allows us to distinguish anticipated acceptance (a pre-use construct) from confirmed acceptance (a post-use construct), and to interpret directional changes in PU, BI, and trust as evidence of positive or negative disconfirmation rather than as undifferentiated change.
(2) By applying k‑means, we move beyond aggregate averages to identify distinct user groups with respect to acceptance. We show that within a single DOT, there are multiple adoption personas, each with different levels of perceived usefulness, ease, and trust. Furthermore, by examining how individuals transition between these groups after use and training, we shed light on the migration of adoption readiness. These findings bridge a gap in understanding how workforce diversity in skills, attitudes, and experiences influences AI tool uptake. (3) Finally, we translate our findings into practical recommendations for public-sector AI implementation. The insights into which factors change significantly with training and which remain low for certain clusters can help DOT leaders tailor their adoption strategies. 
In summary, by measuring and analyzing adoption through a behavioral science lens, this research offers actionable insights to guide responsible and effective AI integration in transportation organizations.

The remainder of this paper is organized as follows: first, we describe the survey methodology and analytical approach, including the TAM construct measurements and clustering technique. We then present the results, detailing changes in acceptance metrics over time and the identified user clusters and their transitions. Finally, we discuss the implications for technology acceptance theory and provide recommendations for practitioners in public agencies seeking to implement generative AI tools at scale. 

\section{Method}
\subsection{Organizational Setting and Participants}
This study was conducted within the North Carolina Department of Transportation (NCDOT), a public agency with approximately 11{,}000 employees spanning planning, engineering, maintenance, information technology, and administrative functions. In 2025, NCDOT launched an eight-week pilot of Microsoft 365 Copilot involving approximately 175 employees drawn from multiple business units, allowing the pilot to capture perceptions from employees with varied administrative, technical, planning, and program-management responsibilities.

\subsection{Pilot Intervention and Governance Conditions}
After completing the pre-pilot survey, participants completed a mandatory Learning Management System (LMS) prerequisite before receiving access to Microsoft 365 Copilot. The LMS prerequisite required approximately 1-2 hours and introduced participants to: (1) fundamentals of generative AI and large language models, (2) prompting strategies, (3) data privacy and security considerations specific to NCDOT systems, and (4) organizational policies governing acceptable use. Participants were required to complete this prerequisite before system access was enabled.

Following the prerequisite, participants received access to Copilot within the standard NCDOT Microsoft 365 environment, including Word, Outlook, Teams, PowerPoint, Excel, and OneNote. Access was limited to organizational data and applications within NCDOT-approved Microsoft 365 environments; use with external or non-approved data sources was restricted. The eight-week pilot combined structured training with hands-on use in routine work tasks. Training was delivered through live virtual sessions, recorded tutorials, self-paced materials, office hours, and weekly FAQ communications. Weekly training emphasized specific applied use cases, including drafting documents in Word, summarizing email threads in Outlook, generating meeting notes in Teams, creating presentations in PowerPoint, and supporting information retrieval.

The pilot was conducted under defined responsible use guardrails. Participants were instructed to avoid sensitive or high risk uses, including handling confidential or personally identifiable information, generating external communications without human review, or using Copilot to make policy, legal, engineering, or operational decisions without validation. Low risk internal tasks, such as drafting internal communications and summarizing non-sensitive content, were encouraged. All AI-generated outputs remained subject to human review before use. These guardrails were communicated through the LMS prerequisite, training sessions, office hours, FAQ emails, and pilot-support materials.

\subsection{Study Design and Survey Timing}
\begin{figure}[H]
\centering
\includegraphics[width=0.97\columnwidth]{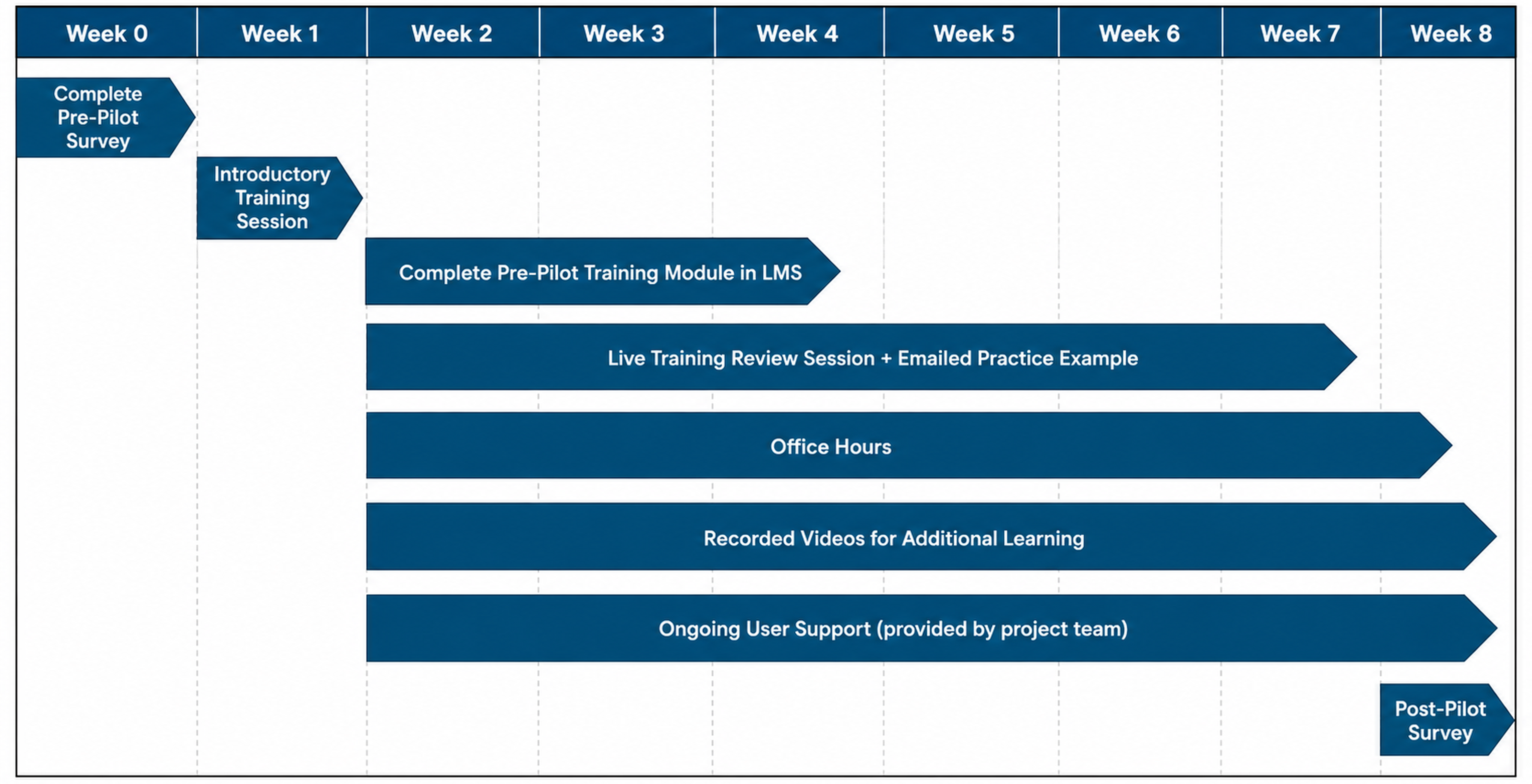}
\caption{Study timeline showing pre-survey (Week 0), training and Copilot usage period (Weeks 1–8), and post-survey (Week 8).}
\label{fig:timeline}
\end{figure}

The study used a two-wave matched-panel design to compare employee perceptions before and after exposure to Copilot. The pre-pilot survey was administered in Week 0, before participants completed pilot training or received Copilot access. The post-pilot survey was administered during Week 8, after participants had completed the LMS prerequisite, participated in ongoing training activities, and had several weeks of opportunity to use Copilot in routine work tasks. Figure~\ref{fig:timeline} provides an overview of the study timeline.

\begin{itemize}
    \item \textbf{Week 0:} Pre-pilot survey administered before training and system access.
    \item \textbf{Weeks 1–8:} Participants completed training activities and used Copilot in routine work tasks under defined responsible-use guardrails.
    \item \textbf{Week 8:} Post-pilot survey administered during the final week of the pilot.
\end{itemize}

\subsection{Survey Instrument and Measures}
Both survey waves contained 16 items, with four items each measuring Perceived Usefulness (PU), Perceived Ease of Use (PEOU), Behavioral Intention (BI), and Trust (TR). All items were rated on the same five-point Likert scale (1 = strongly disagree, 2 = disagree, 3 = neutral, 4 = agree, 5 = strongly agree), with higher scores indicating more favorable perceptions. Item wording was adapted from established Technology Acceptance Model (TAM) and AI-trust measures as shown in Table~\ref{tab:constructs_items_alpha}.
The post-pilot instrument retained the same construct structure and item content but used minimally revised wording to reflect experienced use rather than anticipated use. Accordingly, the two waves should be interpreted as parallel measures of the same four constructs rather than as strictly identical item forms. The phrasing of the post-implementation survey items was slightly modified to reflect past-oriented experiences, in contrast to the future-oriented expectations captured in the pre-pilot survey. The items shown in Table \ref{tab:constructs_items_alpha} reflect the phrasing of the pre-pilot survey. 

\subsection{Data Preparation}
The pilot enrolled 175 employees, all of whom completed the pre-pilot survey as a condition of participation. The pre-pilot wave therefore constitutes a census of the pilot population rather than a sample. Of these 175 participants, 133 also completed the post-pilot survey, so the 76\% figure reflects two-wave completion across the full enrolled cohort rather than a conventional survey response rate.
Responses were de-identified and matched across waves using anonymous unique identifiers. Longitudinal analyses were restricted to respondents who completed both survey waves, resulting in 133 matched cases, or 76\% of the pre-pilot response pool. To assess potential attrition bias, baseline construct scores were compared between respondents who completed both survey waves and respondents who completed only the pre-pilot survey. Because the baseline construct distributions were non-normal, Mann–Whitney U tests were used, with rank-biserial correlations reported as effect sizes. To reduce the influence of inattentive responding, we screened for straight-lining by calculating the within-respondent standard deviation across the 16 items in each survey wave. Respondents with zero within-person variance in either the pre-pilot or post-pilot survey were excluded, yielding a final analytic sample of 124 matched respondents. 

All analyses reported below are based on this final matched sample unless otherwise noted.
After matching and screening, item coding was checked to ensure that higher values consistently represented more favorable perceptions in both waves. Internal consistency was evaluated separately for each construct and survey wave using Cronbach's alpha. Because reliability was acceptable for all four constructs, item responses within each construct were averaged to create composite PU, PEOU, BI, and TR scores for subsequent analyses. 

\subsection{Statistical Analysis of Construct-Level Changes}
Descriptive statistics were computed for participant characteristics, prior technology use, and the four acceptance constructs. For each construct, pre/post changes were evaluated using paired comparisons among matched respondents. Normality of pre/post difference scores was assessed using the Shapiro–Wilk test. Because difference scores deviated from normality, Wilcoxon signed-rank tests were used to assess changes in PU, PEOU, BI, and TR between the pre- and post-pilot surveys. Effect sizes were reported as r, calculated as the standardized test statistic divided by the square root of the number of paired observations. Where multiple related tests were conducted, p-values were adjusted using the Benjamini–Hochberg false-discovery-rate procedure. Statistical significance was evaluated using an adjusted alpha level of 0.05.

\subsection{User Clustering and Longitudinal Transition Analysis}
We used the four construct level composites (PU, PEOU, BI, and TR) to identify baseline acceptance personas and then evaluate how individual participants moved relative to those personas after the pilot. The analysis proceeded in four stages: (1) feature standardization, (2) selection of the number of clusters, (3) estimation of a pre-pilot clustering model and assignment of post-pilot cases to fixed pre-pilot centroids, and (4) computation of transition patterns and construct level changes.

\subsubsection{Step 1. Feature standardization}
Prior to clustering, the pre-pilot survey features were standardized using z-score normalization so that each construct contributed equally to distance-based clustering. Standardization parameters were estimated using only the pre-pilot data. The same pre-pilot means and standard deviations were then applied to transform post-pilot construct scores, ensuring that post-pilot observations were evaluated in the same feature space as the baseline personas.

\subsubsection{Step 2. Selecting the number of clusters}
To identify the most appropriate number of clusters for segmenting user profiles, we applied three established methods: the Elbow Method, the Silhouette Score, and the Calinski–Harabasz (CH) Index. Each method provides a different perspective on clustering performance, allowing for a more robust and data-driven selection of the optimal number of clusters.

The elbow method evaluates the within-cluster sum of squares (WCSS) as a function of the number of clusters. WCSS quantifies the total distance between each data point and the centroid of its assigned cluster, where lower WCSS values indicate tighter and more cohesive groupings. Mathematically, WCSS is defined as:

\begin{equation}
\text{WCSS} = \sum_{i=1}^{k} \sum_{x \in C_i} \|x - \mu_i\|^2
\end{equation}
\\
where,
$k$ is the number of clusters,
$C_i$ is the set of data points in cluster $i$,
$\mu_i$ is the centroid of cluster $i$,
$x$ is a data point in cluster $C_i$,
$\|\cdot\|$ denotes the Euclidean norm.
As the number of clusters increases, WCSS naturally decreases since more centroids allow data points to be grouped into smaller, more compact clusters. This method involves plotting WCSS against different values of $k$ to identify the elbow point, where the rate of decrease in WCSS sharply changes. This point represents a balance between minimizing intra-cluster distance and avoiding overfitting due to excessive partitioning.

The Silhouette Score provides a measure of how well-separated and cohesive the clusters are by quantifying the similarity of each data point to its own cluster compared to other clusters. The scores range from $-1$ to $1$, with higher values indicating more well-defined and distinct clusters. We computed the average silhouette score for a range of cluster counts to evaluate the quality of different clustering solutions. This Score, $s(i)$, for a data point $i$ is defined as:

\begin{equation}
s(i) = \frac{b(i) - a(i)}{\max \{ a(i), b(i) \}}
\end{equation}
\\
where,
$a(i)$ is the average distance between point $i$ and all other points in the same cluster,
$b(i)$ is the minimum average distance between point $i$ and all points in any other cluster (i.e., the nearest neighboring cluster that $i$ is not a part of).

The Calinski–Harabasz Index, also known as the variance ratio criterion, evaluates clustering performance by comparing the between-cluster dispersion with the within-cluster dispersion. A higher index value indicates more compact and well-separated clusters. In this study, CH scores were computed for cluster counts ranging from 2 to 9 to help determine the most suitable number of clusters. The Calinski–Harabasz Index (CH) is calculated as:

\begin{equation}
\text{CH} = \frac{\text{Between-cluster variance} / (k - 1)}{\text{Within-cluster variance} / (n - k)}
\end{equation}
\\
where,
$k$ is the number of clusters,
$n$ is the total number of data points.

\subsubsection{Step 3. Pre/post clustering and migration tracking} 
The standardized pre-pilot survey vectors of four constructs (PU, PEOU, BI, and TR) were partitioned using a K-Means clustering algorithm. 
K-means was selected over Partitioning Around Medoids (PAM) or Gaussian mixtures models (GMM) due to its reliance on Euclidean distance metric, which allowed us to fix the centroids that characterize each cluster in the pre-pilot survey phase and then consistently apply them to assign participants to clusters in the post-survey phase, following prior work that groups transportation assets using unsupervised learning techniques \cite{Karimzadeh2021Optimal, bansal2017improved, kanungo2002efficient}. 

To analyze changes in perception following the training and use of the technology, the same four metrics from the post-survey (PU, PEOU, BI, and TR) were extracted and standardized using the same z-score scaling parameters derived from the pre-pilot survey data to maintain consistency in feature space. These transformed post-survey responses were then fed into the previously trained K-Means model to determine each participant's post-survey cluster label. This approach enabled a direct comparison of cluster membership transitions from pre- to post-pilot survey, offering insight into how participants' perceptions evolved.

\subsubsection{Step 4. Cluster Membership Transitions}
To evaluate how participants’ acceptance profiles changed following the pilot, we compared each respondent’s pre-pilot cluster assignment with their post-pilot assignment to the fixed baseline centroids. The resulting transition matrix captured the number and percentage of participants who remained in the same persona or moved to another persona. Percentages were calculated relative to the number of participants in each pre-pilot cluster. For each transition path, we also computed mean changes in PU, PEOU, BI, and TR to describe the construct-level shifts associated with upward, downward, or stable persona movement.

As a robustness check, Gaussian mixture models were estimated using the same standardized pre-pilot features. Competing two-, three-, and four-profile solutions were compared across alternative covariance structures using Akaike Information Criterion and Bayesian Information Criterion. These models were used to assess whether the selected k-means solution was broadly consistent with alternative person-centered segmentation approaches.

All analyses were conducted in Python 3.12.4. The Python analytic stack comprised pandas (2.2.2) and NumPy (1.26.4) for data handling; SciPy (1.13.1), statsmodels (0.14.2), and Pingouin (0.6.0) for statistical testing; scikit-learn (1.4.2) for clustering; and Matplotlib (3.8.4) and Plotly 
(6.7.0) for visualization. The k-means model was estimated using 20 random initializations  and a fixed random seed of 42 to support reproducibility. All analysis scripts and version specifications are available from the corresponding author upon request.

\subsection{Secondary Outcomes and Transition-Group Comparisons}
In addition to the four TAM-trust constructs, the survey collected secondary measures related to anticipated and actual Copilot use cases, user concerns, perceived time savings, perceived work impacts, perceived need for customized AI tools, and perceived productivity impact in the absence of Copilot. Multi-select task-use items were grouped into six categories: data/charts, information retrieval, brainstorming, communications, presentations, and summarization. Concern items were grouped into accuracy, privacy/data, job/skills, and no-concern categories. These variables were coded as binary indicators at the participant level.

For interpretability, cluster-transition paths were also grouped into two broader categories. Participants who remained in the same persona or moved to a more favorable persona were classified as having a stable or improved transition. Participants who moved to a less favorable persona were classified as having a declined transition. These terms (\textit{stable/improved} and \textit{declined}) are used throughout the remainder of the paper to distinguish between participants whose acceptance profile stayed the same or improved and those whose acceptance profile shifted downward. This grouping was used only for secondary descriptive and comparative analyses and the primary transition analysis retained the full cluster-to-cluster migration matrix.

Within-group pre/post changes in binary task-use and concern indicators were assessed using McNemar’s test. Between-group differences in change patterns were assessed using chi-square tests or Fisher’s exact tests when expected cell counts were small. Continuous or ordinal post-pilot outcomes, including perceived time savings and work-impact ratings, were compared between transition groups using Welch’s t-test or Mann–Whitney U tests depending on distributional assumptions. For categorical post-pilot outcomes, chi-square tests of independence were used, with Cramer’s V reported as an effect size. P-values were adjusted within related families of tests using the Benjamini–Hochberg procedure.

\subsection{Qualitative Analysis}
To complement the quantitative analysis and provide deeper insight into participants' perceptions, an exploratory keyword-based content analysis was conducted on open-ended survey responses collected before and after the Copilot pilot. The analysis aimed to map participants' qualitative feedback to key constructs of the Technology Acceptance Model (TAM), including PU, PEOU, TR, and BI.

Qualitative responses were extracted from multiple open-ended survey items, including pre-pilot questions on concerns and general feedback, as well as post-pilot questions related to experienced benefits, concerns, and overall impressions. Responses were aggregated at the participant level separately for each wave. Standard preprocessing steps were applied, including conversion to lowercase, removal of punctuation, and normalization of whitespace.

A construct-specific keyword dictionary was developed based on TAM and AI-trust literature \cite{Davis1989TAM, venkatesh2000theoretical, venkatesh2003user, gefen2003trust, mcknight2011trust} and refined through review of the observed vocabulary in the responses.
For example, PU was associated with terms such as "improve," "helpful," "useful," "productivity," and "efficient," reflecting performance and efficiency gains. PEOU was captured through terms such as "easy," "simple," "clear," and "understand," indicating usability and learnability. TR included terms such as "accurate," "reliable," "risk," "privacy," "error," and "confidence," representing concerns related to system reliability and data security. BI was identified through terms such as "use," "continue," "plan," "recommend," and "future," reflecting participants’ intentions and willingness to adopt the system.

Then two complementary metrics were used to quantify construct representation in the qualitative data. The first was a presence-based metric indicating whether a participant used at least one keyword associated with a construct. The second was a frequency-based metric capturing the number of construct-related keyword occurrences per participant. These metrics were summarized separately for the stable/improved and declined transition groups.
Keyword matching was performed using exact word-boundary searches to minimize false positives and ensure consistent coding across responses.

To check coding validity, a random subset of 12 responses was manually reviewed to assess whether keyword matches were contextually consistent with the intended construct. Disagreements or ambiguous terms were resolved through discussion and used to refine the dictionary before final coding.

\section{Results}
\subsection{Attrition bias}
To examine potential attrition bias, we compared participants who completed both survey waves with those who completed only the pre-pilot survey. Baseline scores for perceived usefulness, perceived ease of use, behavioral intention, and trust were compared to assess whether retained and non-retained participants differed before pilot exposure.
Normality assumptions were evaluated using the Shapiro--Wilk test, which indicated non-normal distributions ($p < 0.05$). Therefore, the Mann--Whitney U test was used to compare retained and dropout participants. Effect sizes were reported using rank-biserial correlation.

The results, summarized in Table~\ref{tab:attrition}, indicate that there were no statistically significant differences between retained and dropout participants across all constructs ($p > 0.05$). The observed differences in mean scores were small, and the corresponding effect sizes were negligible, suggesting minimal practical differences between the two groups and participant attrition was largely random rather than systematic. These results provide no evidence of systematic attrition bias with respect to baseline TAM and trust measures.

\begin{table}[H]
\centering
\small
\caption{Attrition analysis comparing baseline construct scores between retained and non-retained participants.}
\label{tab:attrition}
\begin{tabular}{lccccc}
\toprule
\multirow{2}{*}{Construct} 
& Retained & Non-retained & Mean difference & Mann--Whitney & Effect size \\
& ($n=133$) & ($n=42$) & (Retained $-$ Non-retained) & ($p$) & ($r_{\mathrm{rb}}$) \\
\midrule
PU   & 3.82 $\pm$ 0.71 & 3.97 $\pm$ 0.62 & -0.15 & 0.310 & -0.10 \\
PEOU & 3.57 $\pm$ 0.66 & 3.68 $\pm$ 0.67 & -0.11 & 0.462 & -0.07 \\
BI   & 3.76 $\pm$ 0.65 & 3.71 $\pm$ 0.56 &  0.05 & 0.444 &  0.08 \\
TR   & 3.20 $\pm$ 0.75 & 3.33 $\pm$ 0.60 & -0.13 & 0.373 & -0.09 \\
\bottomrule
\multicolumn{6}{p{0.95\linewidth}}{\footnotesize \textit{Note.} Values are mean $\pm$ standard deviation. $r_{\mathrm{rb}}$ denotes the rank-biserial correlation.} \\
\end{tabular}
\end{table}

\subsection{Descriptive Statistics}
Table \ref{tab:sample-descriptives} provides an overview of the 124 respondents included in the analysis. The majority of participants (90\%) are non-executives. Overall, the group is relatively experienced, with 75\% having more than three years of professional experience. 
The experience levels are fairly evenly distributed across early-career, mid-career, and late-career groups, suggesting a generally seasoned workforce primarily composed of experienced knowledge workers.

\begin{table}[H]
  \small
  \centering
  \caption{Descriptive characteristics of the final analytic sample (N = 124)}
  \label{tab:sample-descriptives}
  \renewcommand{\arraystretch}{1}
  \setlength{\tabcolsep}{6pt}
  \begin{tabular}{llrr}
    \toprule
    \textbf{Variable} & \textbf{Category} & \textbf{Frequency} & \textbf{\% of Sample} \\
    \midrule
    \multirow{3}{*}{Executive status} 
        & No  & 112 & 90.3 \\
        & Yes &  12 &  9.7 \\
        & \textit{Missing} & 0 & 0.0 \\
    %\midrule
    %\multirow{3}{*}{Team‐management role}
    %    & No  & 47 & 35.3 \\
    %    & Yes & 51 & 38.3 \\
    %    & \textit{Missing} & 35 & 26.3 \\
    \midrule
    \multirow{7}{*}{Years of experience} 
        & $<$1 year        &  7 &  5.6 \\
        & 1–3 years        & 24 & 19.4 \\
        & 4–6 years        & 25 & 20.2 \\
        & 7–10 years       & 21 & 16.9 \\
        & 11–20 years      & 25 & 20.2 \\
        & 20+ years        & 22 & 17.7 \\
        & \textit{Missing} &  0 &  0.0 \\
    \bottomrule
  \end{tabular}
\end{table}

Table~\ref{tab:pre_post_clean} summarizes the aggregate changes in the four survey constructs between the pre-pilot and post-pilot surveys. Overall, the results indicate that participants’ perceptions remained relatively stable across most constructs, with the exception of perceived usefulness, which declined significantly after hands-on experience with Microsoft 365 Copilot. This pattern suggests a recalibration of initial expectations regarding the tool’s usefulness, while perceived ease of use, behavioral intention, and trust showed only small, non-significant changes. Because pre--post difference scores were non-normally distributed, construct-level changes were assessed using Wilcoxon signed-rank tests with Benjamini--Hochberg adjusted p-values.

\begin{table}[H]
\centering
\small
\caption{Pre--post changes in survey construct scores among matched respondents.}
\label{tab:pre_post_clean}
\begin{tabular}{lccccc}
\toprule
\multirow{2}{*}{Construct} 
& Pre-pilot & Post-pilot & Mean change & Wilcoxon & Effect size \\
& ($n=124$) & ($n=124$) & (Post $-$ Pre) & (adjusted $p$) & ($r$) \\
\midrule
PU   & 3.85 $\pm$ 0.71 & 3.62 $\pm$ 0.73 & -0.23 & $<0.001$ & -0.40 \\
PEOU & 3.60 $\pm$ 0.67 & 3.69 $\pm$ 0.63 &  0.09 & 0.136    &  0.21 \\
BI   & 3.80 $\pm$ 0.65 & 3.88 $\pm$ 0.67 &  0.08 & 0.235    &  0.16 \\
TR   & 3.20 $\pm$ 0.77 & 3.24 $\pm$ 0.74 &  0.04 & 0.339    &  0.11 \\
\bottomrule
\multicolumn{6}{p{0.88\linewidth}}{\footnotesize \textit{Note.} Values are mean $\pm$ standard deviation. Mean change represents the post-pilot mean minus the pre-pilot mean.} \\
\end{tabular}
\end{table}

\begin{figure}[H]
    \small
    \centering
    \includegraphics[width=0.99\linewidth]{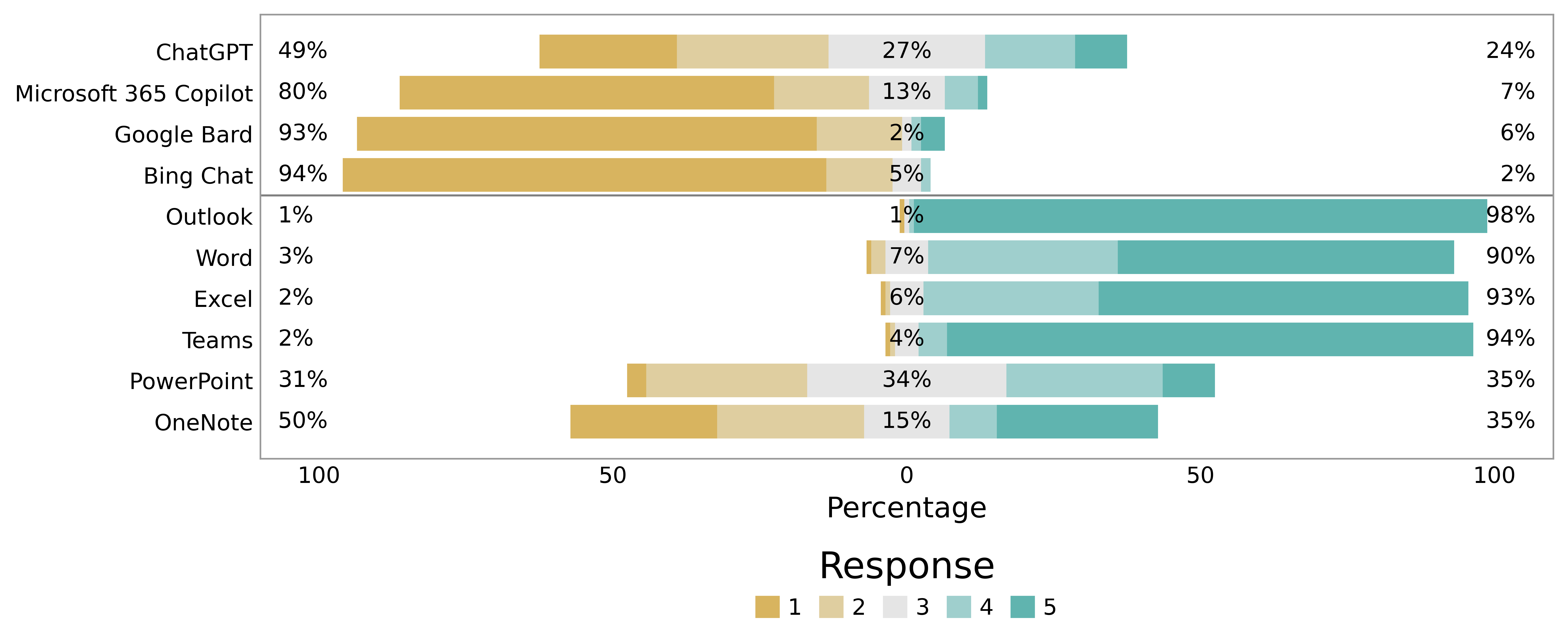}
    \caption{ Baseline frequency of participant-reported use of large language model (LLM) tools and Microsoft 365 applications prior to the pilot (N=124). Responses were recorded on a five-point Likert scale: 1= Never, 2= Rarely, 3= Occasionally, 4= Frequently, and 5= Very Frequently.
    }
    \label{fig:Likert_Pre}
\end{figure}

The baseline frequency of participant-reported use of large language model (LLM) tools and Microsoft 365 applications is shown in Figure~\ref{fig:Likert_Pre}. Respondents first reported their prior use of common LLM-based tools, including ChatGPT, Microsoft 365 Copilot, Google Bard, and Bing Chat. They then reported their typical use of Microsoft 365 applications, including Outlook, Word, Excel, Teams, PowerPoint, and OneNote. Responses were recorded on a five-point scale: 1 = Never, 2 = Rarely (once a month or less), 3 = Occasionally (a few times per month), 4 = Frequently (weekly), and 5 = Very Frequently (daily).
Figure \ref{fig:Likert_Pre} clearly shows limited pre-existing familiarity with LLM tools. Microsoft 365 Copilot, the tool evaluated in this study, was also largely unfamiliar to respondents, with 81\% reporting minimal or no prior use. ChatGPT showed comparatively higher baseline familiarity among the LLM tools. However, approximately half of respondents still reported little to no prior experience with it.

In contrast, participants reported frequent use of core Microsoft 365 applications, particularly Outlook, Word, Excel, and Teams. Use of PowerPoint and OneNote was more varied across respondents. Taken together, these results suggest that the pilot introduced a relatively unfamiliar AI-enabled capability into an otherwise familiar productivity-software environment. This contrast is important for interpreting the adoption results: participants were not learning an entirely new workplace software ecosystem, but they were learning to integrate a new generative AI function into existing Microsoft 365 workflows. This finding reinforces the need for targeted training, workflow specific examples, and change management support during enterprise AI implementation.

\subsection{Construct Reliability Assessment}
Internal consistency reliability was evaluated for each construct using Cronbach's alpha coefficients ($\alpha$) for both pre- and post-pilot survey data. As shown in Table~\ref{tab:constructs_items_alpha}, all constructs exceeded the commonly used threshold of 0.70 in both survey waves, indicating acceptable to high internal consistency. These results support the use of averaged construct-level scores for perceived usefulness (PU), perceived ease of use (PEOU), behavioral intention (BI), and trust (TR) in the subsequent analyses.

\begin{table}[H]
  \footnotesize
  \centering
  \caption{Survey Constructs, Questions, and Scale Reliabilities}
  \label{tab:constructs_items_alpha}
  \renewcommand{\arraystretch}{1.4}
  \begin{tabularx}{\textwidth}{%
      >{\raggedright\arraybackslash}l  
      >{\raggedright\arraybackslash}l  
      >{\raggedright\arraybackslash}X  
      >{\centering\arraybackslash}c  
      >{\centering\arraybackslash}c  
    }
    \toprule
    \textbf{Construct}
      & \textbf{Code}
      & \textbf{Survey Item}
      & \textbf{\shortstack{$\alpha$\\(Pre)}} 
      & \textbf{\shortstack{$\alpha$\\(Post)}} \\
    \midrule

    PU: Perceived Usefulness
      & PU1
      & Using Microsoft 365 Copilot will improve my job performance.
      & 0.942
      & 0.926 \\
      & PU2
      & Microsoft 365 Copilot will help me accomplish tasks more quickly.
      & & \\
      & PU3
      & Microsoft 365 Copilot will increase my productivity on work-related tasks.
      & & \\
      & PU4
      & Overall, I think Microsoft 365 Copilot will be beneficial for my job.
      & & \\

    \midrule
    PEOU: Perceived Ease of Use
      & PEOU1
      & Learning to operate Microsoft 365 Copilot will be easy for me.
      & 0.924
      & 0.882 \\
      & PEOU2
      & Interacting with Microsoft 365 Copilot will be clear and understandable.
      & & \\
      & PEOU3
      & It will be easy to become skillful at using Microsoft 365 Copilot.
      & & \\
      & PEOU4
      & Overall, I think Microsoft 365 Copilot is user-friendly.
      & & \\

    \midrule
    BI: Behavioral Intention
      & BI1
      & I intend to use Microsoft 365 Copilot regularly for my work.
      & 0.852
      & 0.865 \\
      & BI2
      & I predict I will use Microsoft 365 Copilot whenever it can assist me.
      & & \\
      & BI3
      & I plan to increase my use of Microsoft 365 Copilot in the near future.
      & & \\
      & BI4
      & I would recommend Microsoft 365 Copilot to my colleagues.
      & & \\

    \midrule
    TR: Trust
      & TR1
      & I believe Microsoft 365 Copilot will provide reliable outputs or recommendations.
      & 0.930
      & 0.897 \\
      & TR2
      & I think I will be able to depend on Microsoft 365 Copilot to function correctly at critical moments.
      & & \\
      & TR3
      & I trust Microsoft 365 Copilot's ability to handle sensitive or essential data.
      & & \\
      & TR4
      & Overall, I feel confident about Microsoft 365 Copilot's performance and integrity.
      & & \\

    \bottomrule
  \end{tabularx}

  \begin{tablenotes}[flushleft]\footnotesize
    \item \textit{Note.} Reliability estimates are Cronbach's $\alpha$ for each construct (pre-pilot vs.\ post-pilot).
  \end{tablenotes}
\end{table}

\subsection{User Clusters and Pre–Post Perception Dynamics}
To identify baseline acceptance personas, we evaluated candidate k-means solutions using the elbow method, average silhouette score, Calinski--Harabasz (CH) index, cluster-size balance, and practical interpretability. The goal of this analysis was not only to maximize statistical separation, but also to identify user groups that were sufficiently distinct and meaningful for implementation planning.

\begin{figure}[H]
    \centering
    \includegraphics[width=0.5\textwidth]{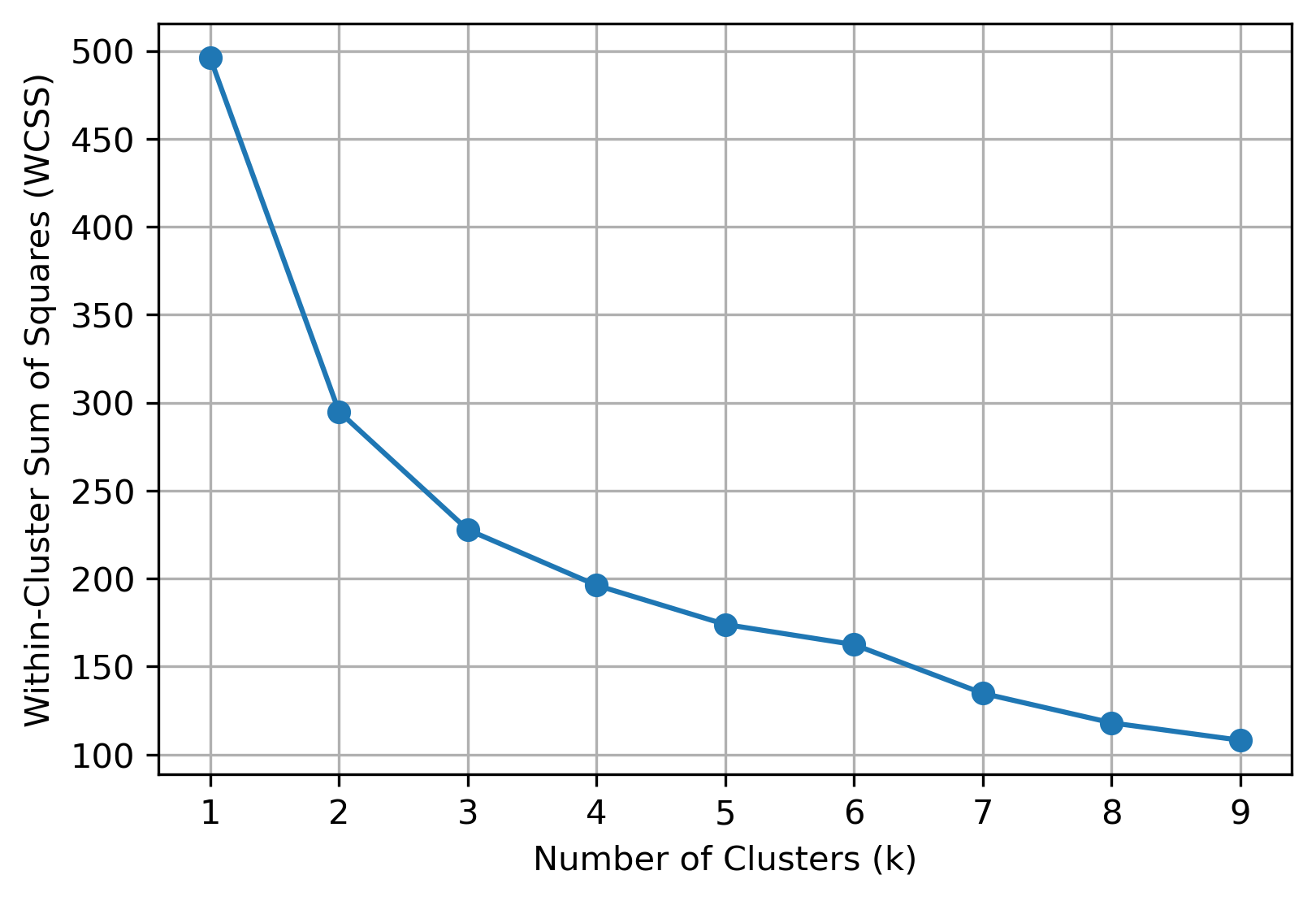}
    \caption{Elbow Method for evaluating the optimal number of clusters.}
    \label{fig:elbow}
\end{figure}

The elbow method, shown in Figure~\ref{fig:elbow}, indicated a clear reduction in within-cluster sum of squares (WCSS) as the number of clusters increased. However, the rate of improvement began to diminish after $k=3$, suggesting limited additional explanatory value from more complex cluster solutions.

The silhouette and CH diagnostics favored a more parsimonious two-cluster solution. However, further examination showed that the two-cluster solution was highly imbalanced, with one cluster containing nearly all respondents and the other capturing only a small group of extreme cases. This structure was less useful for implementation-oriented workforce segmentation. In contrast, the three-cluster solution provided a more interpretable and practically meaningful structure while still maintaining adequate separation among participant profiles. Therefore, the three-cluster solution was retained for subsequent persona development and transition analysis.

\begin{figure}[H]
    \centering
    \begin{subfigure}[b]{0.49\textwidth}
        \includegraphics[width=\textwidth]{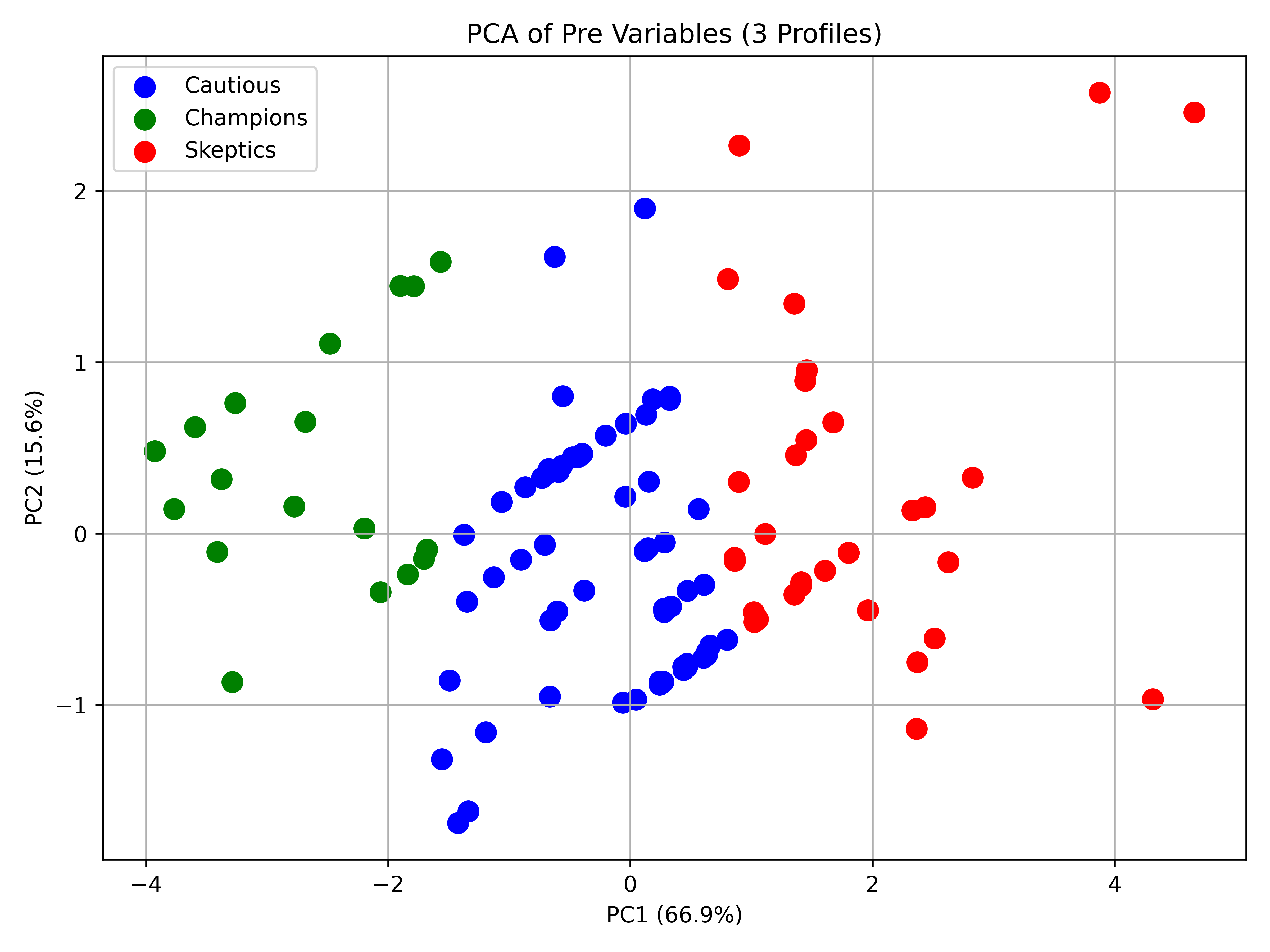}
        \caption{Three-cluster solution}
    \end{subfigure}
    \hfill
    \begin{subfigure}[b]{0.49\textwidth}
        \includegraphics[width=\textwidth]{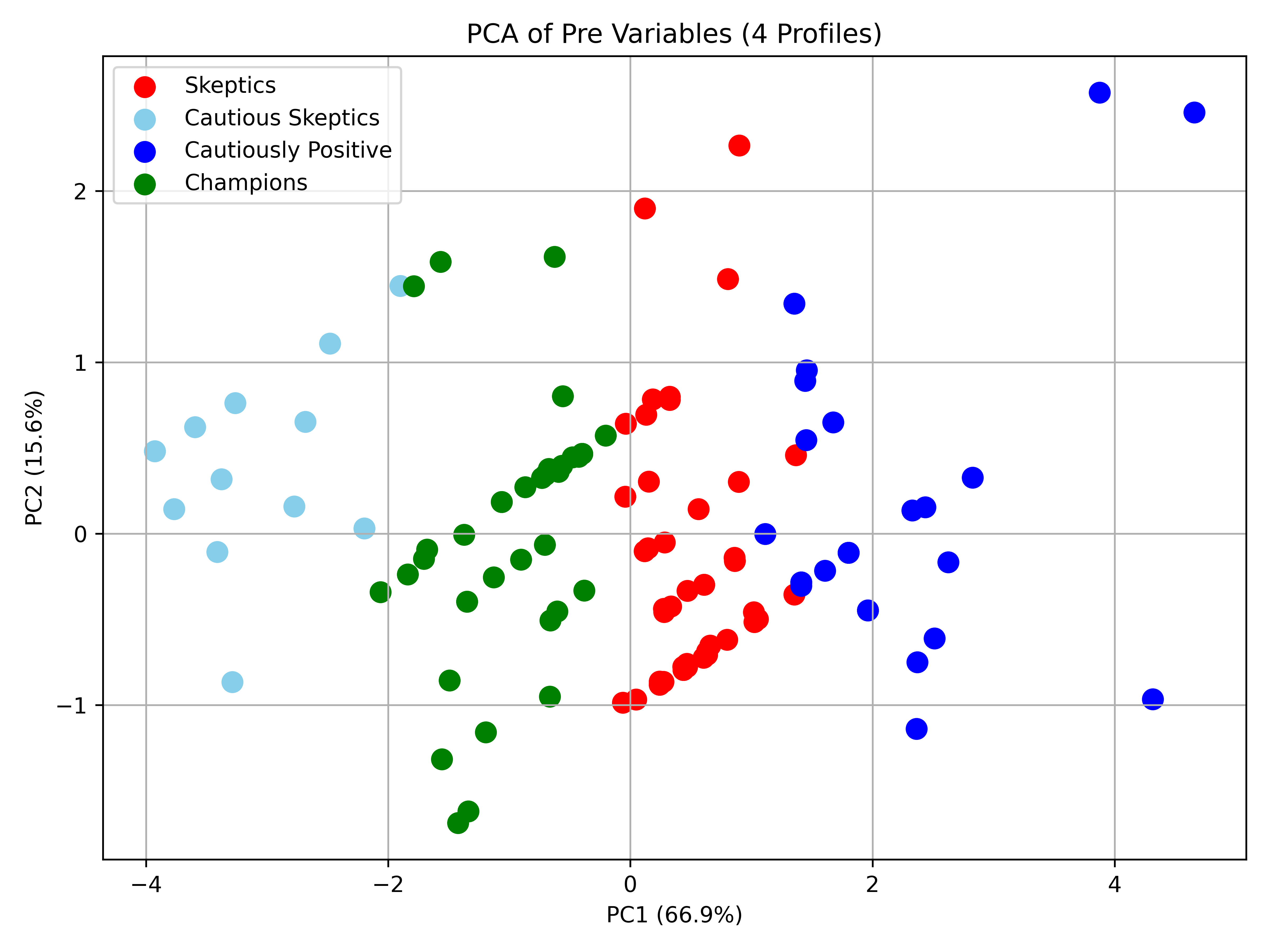}
        \caption{Four-cluster solution}
    \end{subfigure}
    \caption{PCA visualization of candidate k-means cluster solutions. (a) selected three-cluster solution, which produced the most interpretable structure for identifying baseline acceptance personas. (b) four-cluster solution for comparison.}
    \label{fig:pca_profiles}
\end{figure}

To further assess this selection, Principal Component Analysis (PCA) was used as a visualization tool to examine cluster separation in a reduced two-dimensional space. As shown in Figure~\ref{fig:pca_profiles}, the three-profile solution exhibits clearer separation and more coherent grouping compared to the four-profile solution, which shows increased overlap among intermediate profiles. Importantly, PCA was used solely for visualization purposes and not for determining the optimal number of clusters.

Based on the selected three-cluster solution, participants were categorized into three personas reflecting varying levels of perception across PU, PEOU, BI, and TR. As also summarized in Table \ref{tab:pre_cluster_summary}, the three clusters were interpreted and labeled as follows:

\begin{itemize}
    \item \textit{Cluster 0 - Skeptics (n=35, 28\%)}: Members of this group
    %begin the pilot from a position of caution. They 
    score well below the overall mean on every construct. Their Trust score (2.61) is below the sample mean, and their Perceived Usefulness (3.05) is the lowest of any group. Also,  their score for Ease of Use (3.18) remains only slightly above neutral.
   
    \item \textit{Cluster 1 - Cautiously Positive (n=70, 56\%)}: This cluster, comprising over half of the pilot group, reflected mid-range evaluations with scores near the agree threshold (PU = 4.01, BI = 3.91), indicating a more balanced and cautiously positive outlook toward the LLM tool. 
    %They appear open to experimentation but still require evidence, through training or peer endorsements, before fully committing.
    \item \textit{Cluster 2 - Champions (n=19, 15\%)}: Members of this cluster rate every dimension very highly. They strongly agree that Copilot will help them (PU=4.75) and are confident they can master it (PEOU=4.61). Their Trust score (4.22) is a full point higher than the overall average, and their Behavioral Intention (4.66) indicates they already plan to integrate Copilot into routine tasks. 
    %signalling enthusiasm and readiness to champion the tool among peers.
\end{itemize}

\begin{table}[H]
\small
\centering
\caption{Average pre-pilot construct scores by baseline acceptance persona.}
\label{tab:pre_cluster_summary}
\renewcommand{\arraystretch}{1.15}
\begin{tabular}{lccccc}
\toprule
\textbf{Cluster} & \textbf{n} & \textbf{PU} & \textbf{PEOU} & \textbf{BI} & \textbf{TR} \\
\midrule
Skeptics (C0) & 35 & 3.05 & 3.18 & 3.09 & 2.61 \\
Cautiously Positive (C1) & 70 & 4.01 & 3.53 & 3.91 & 3.22 \\
Champions (C2) & 19 & 4.75 & 4.61 & 4.66 & 4.22 \\
\midrule
\textbf{Overall} & 124 & 3.85 & 3.60 & 3.80 & 3.20 \\  % 
\bottomrule
\end{tabular}
\end{table}

\begin{figure}[H]
    \small
    \centering

    % Row 1
    \begin{subfigure}[b]{0.49\textwidth}
        \includegraphics[width=\textwidth]{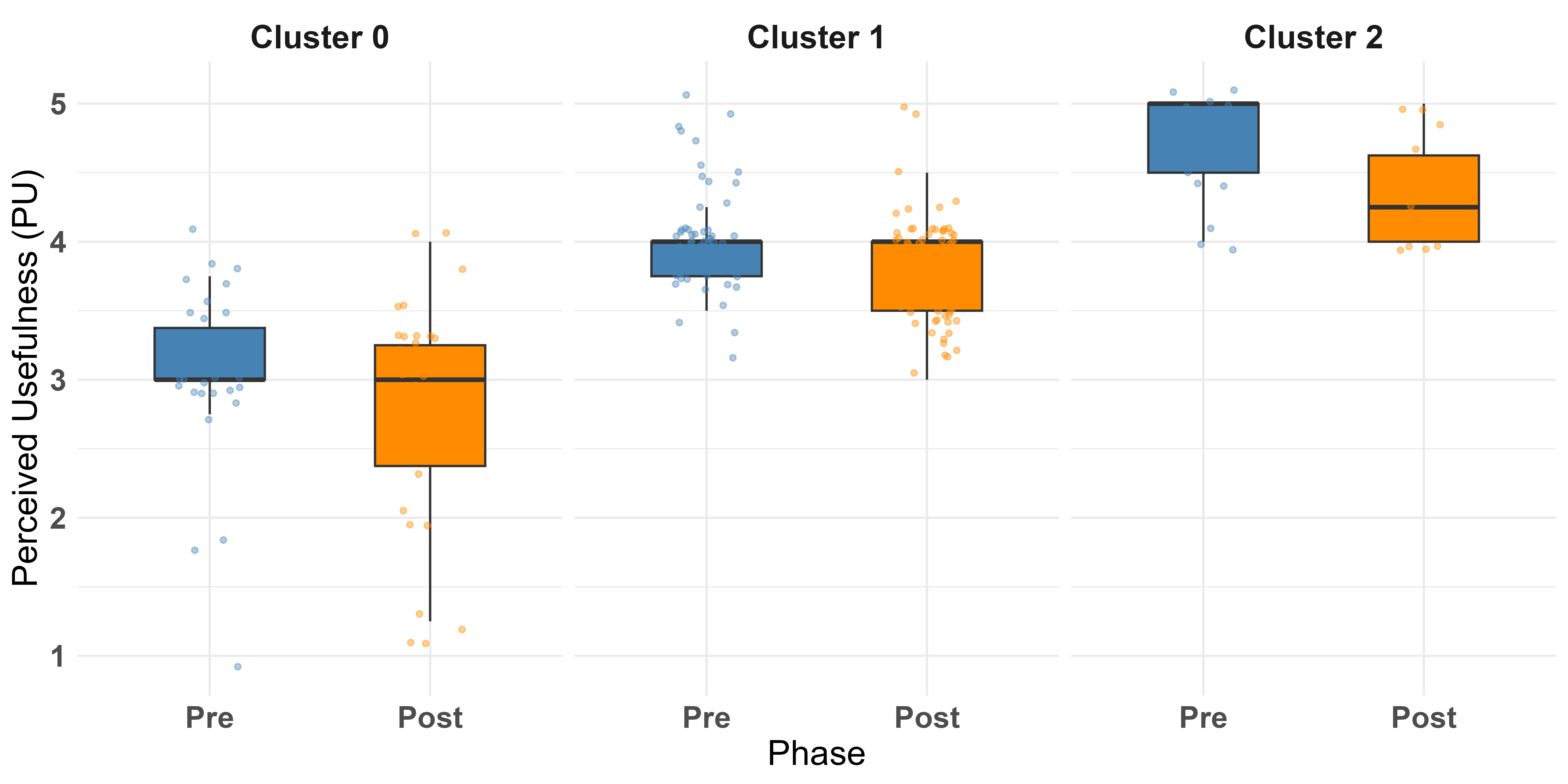}
        \caption{Perceived Usefulness (PU)}
        \label{fig:PU}
    \end{subfigure}
    \hfill
    \begin{subfigure}[b]{0.49\textwidth}
        \includegraphics[width=\textwidth]{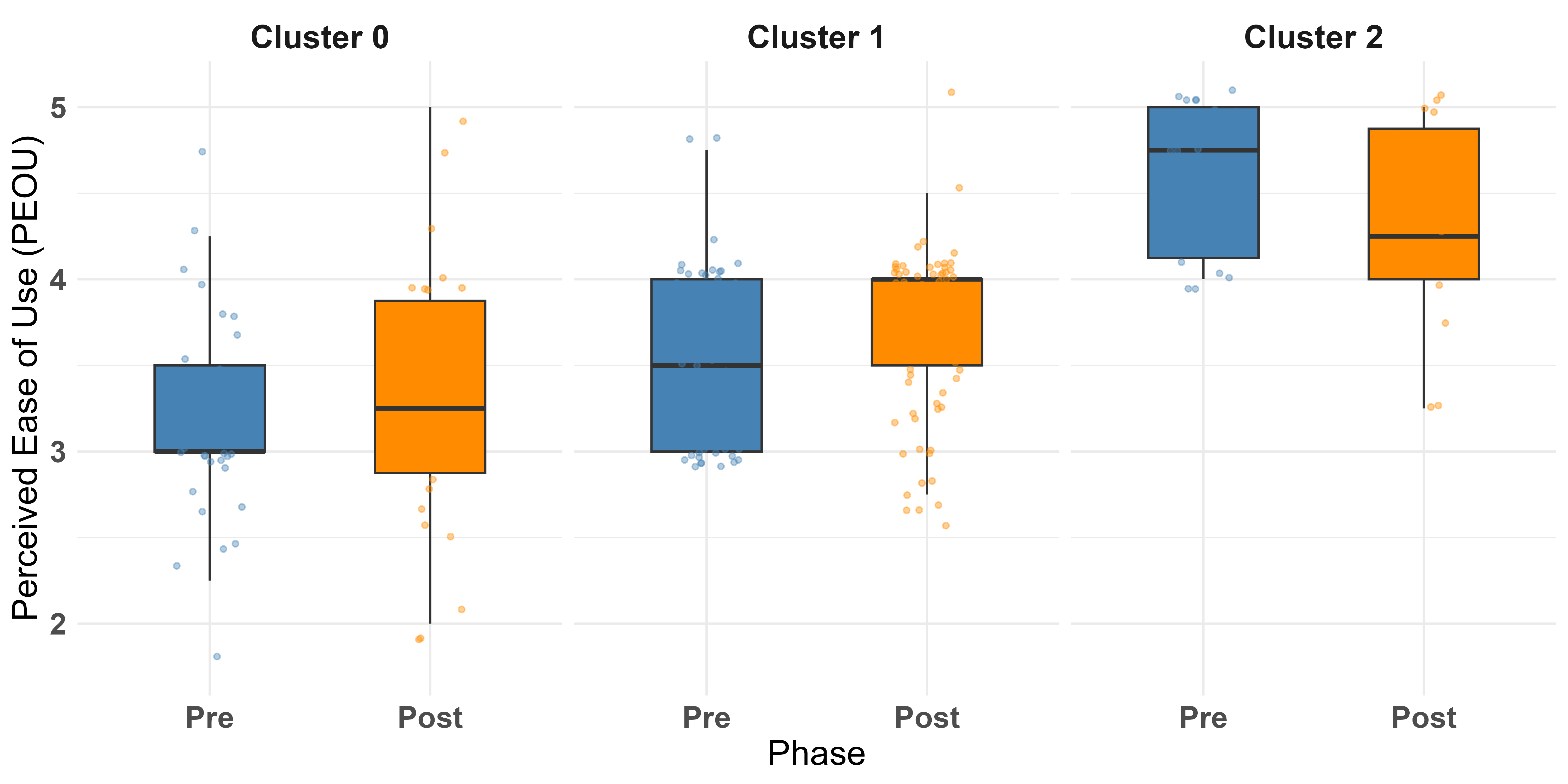}
        \caption{Perceived Ease of Use (PEOU)}
        \label{fig:PEOU}
    \end{subfigure}

    \vspace{0.5cm}

    % Row 2
    \begin{subfigure}[b]{0.49\textwidth}
        \includegraphics[width=\textwidth]{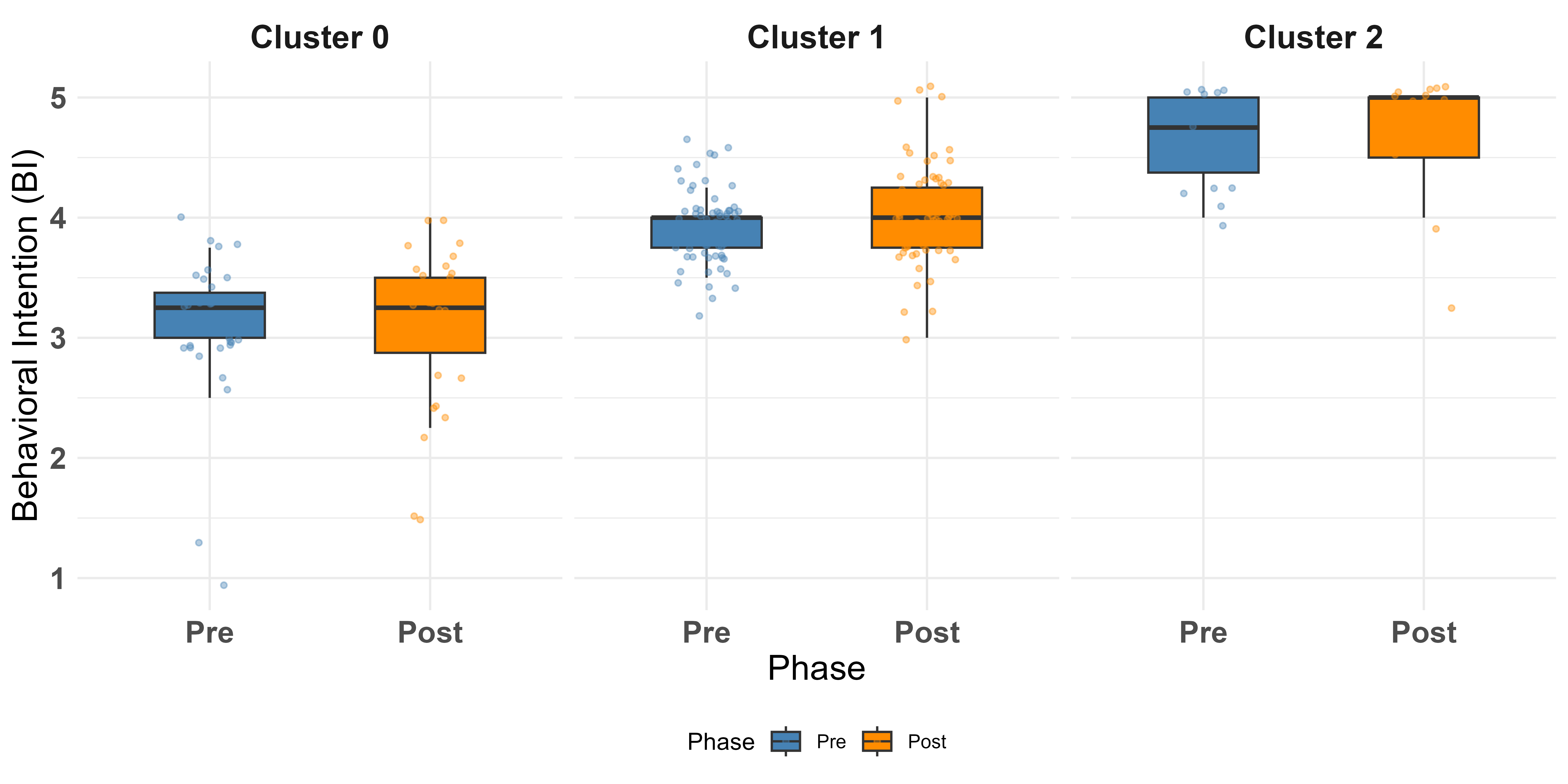}
        \caption{Behavioral Intention (BI)}
        \label{fig:BI}
    \end{subfigure}
    \hfill
    \begin{subfigure}[b]{0.49\textwidth}
        \includegraphics[width=\textwidth]{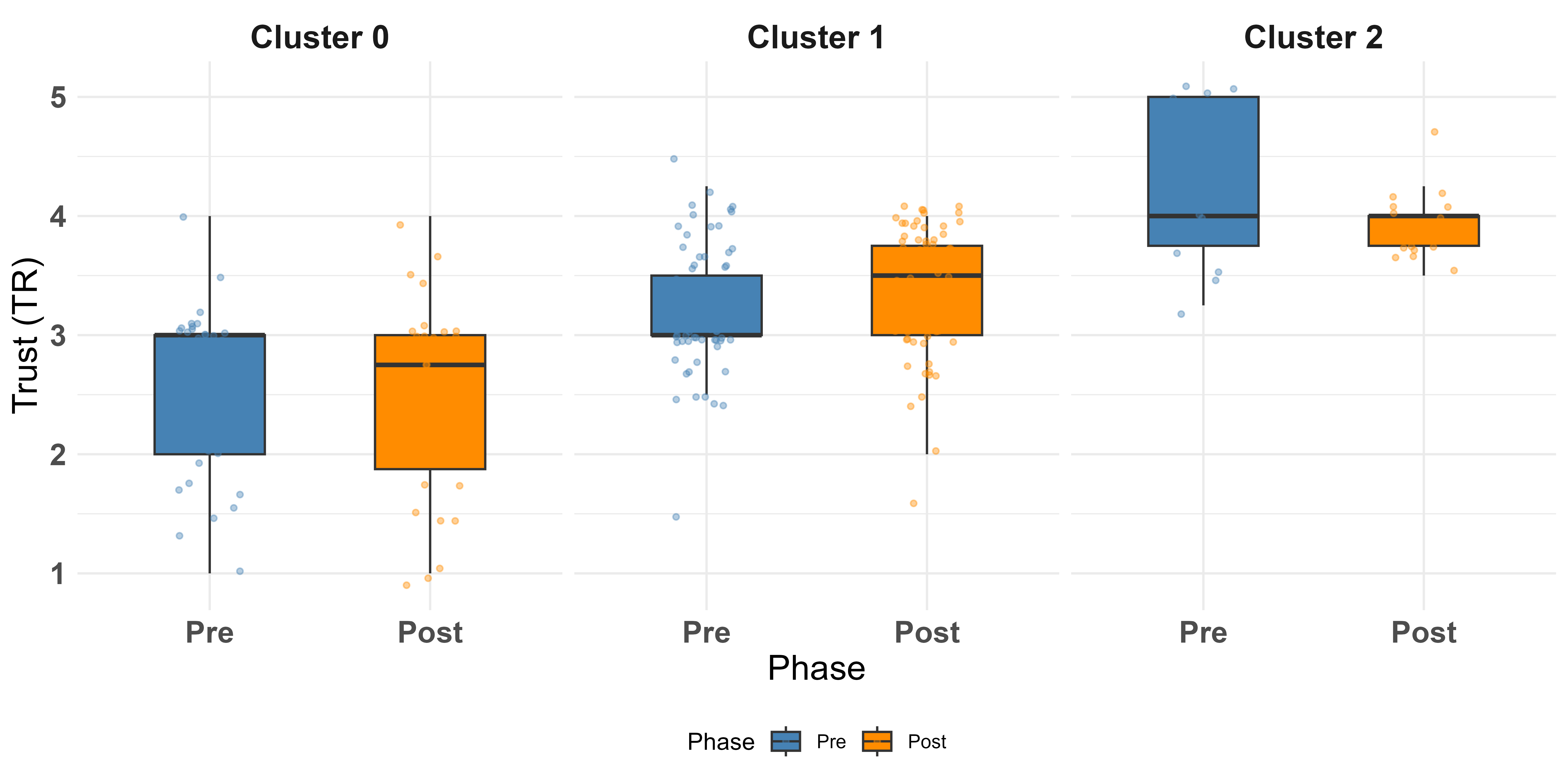}
        \caption{Trust (TR)}
        \label{fig:TR}
    \end{subfigure}

    \caption{Distributions of PU, PEOU, BI, and TR before and after the eight-week pilot for each user cluster (0:Skeptics, 1:Cautiously Positive, 2:Champions)}
    \label{fig:cluster_boxplots}
\end{figure}

To gauge how attitudes shifted within each persona (cluster), we kept the original K-means centroids and reassigned every post-pilot survey response to the same three clusters following the process explained in the method section. Before focusing on the members migrating between clusters, we visualized the pre- and post- distribution of each cluster, using side-by-side boxplots, for each of the four constructs in Figure \ref{fig:cluster_boxplots}.

Table~\ref{tab:between_profile_pre_post} compares mean construct scores across the three baseline acceptance personas at both the pre-pilot and post-pilot survey waves. At baseline, the personas were significantly different across all four constructs. The post-pilot scores show that these between-persona differences remained statistically significant across all constructs after adjustment. Because participants could migrate between personas between survey waves, the pre- and post-pilot persona-level comparisons shown in Table~\ref{tab:between_profile_pre_post} should be interpreted as cross-sectional comparisons within each survey wave rather than repeated-measures comparisons of fixed participant groups.

Pairwise post-hoc comparisons were conducted following the significant Kruskal--Wallis tests to identify which personas differed from one another. The results indicated that all three persona pairs were significantly different across each construct at the pre-pilot stage. This confirms that the three baseline personas represented distinct levels of technology acceptance prior to the pilot. Post-pilot pairwise comparisons showed that the personas also remained distinct across PU, PEOU, BI, and TR.

\begin{table}[H]
\centering
\small
\caption{Between-persona comparisons of constructs at pre-pilot and post-pilot levels.}
\label{tab:between_profile_pre_post}
\resizebox{\textwidth}{!}{%
\begin{tabular}{llccccc}
\toprule
Construct & Score Type & Skeptics & Cautiously Positive & Champions & $H$ & $p_{adj}$ \\
\midrule
PU & Pre  & 3.05 & 4.01 & 4.75 & 84.88 & $<0.001$* \\
   & Post & 3.05 & 3.81 & 3.95 & 33.74 & $<0.001$* \\
\midrule
PEOU & Pre  & 3.18 & 3.53 & 4.61 & 48.77 & $<0.001$* \\
     & Post & 3.54 & 3.64 & 4.13 & 13.71 & 0.001* \\
\midrule
BI & Pre  & 3.09 & 3.91 & 4.66 & 82.69 & $<0.001$* \\
   & Post & 3.36 & 4.07 & 4.17 & 30.20 & $<0.001$* \\
\midrule
TR & Pre  & 2.61 & 3.22 & 4.22 & 50.64 & $<0.001$* \\
   & Post & 2.84 & 3.37 & 3.53 & 10.74 & 0.005* \\
\bottomrule
\end{tabular}%
}
\begin{tablenotes}
\scriptsize
\item Note. Values are reported as mean construct scores. Kruskal--Wallis tests were used for omnibus between-profile comparisons. All $p$-values were adjusted using the Benjamini--Hochberg false discovery rate correction. 
* $p<0.05$.
\end{tablenotes}
\end{table}

\subsection{Transition Matrix and Migration Paths}
Table \ref{tab:cluster_transition} illustrates the transition of participants between clusters from the pre- to the post-pilot phase. It indicates that aggregate persona counts changed only modestly from pre-pilot to post-pilot, but individual-level transitions were substantial. Among baseline Skeptics, 60\% remained Skeptics and 40\% moved to Cautiously Positive. Among baseline Cautiously Positive users, 76\% remained in the same persona, 13\% moved upward to Champions, and 11\% moved downward to Skeptics. Among baseline Champions, only 32\% remained Champions, while 58\% moved to Cautiously Positive and 10\% moved to Skeptics. These patterns indicate convergence toward the Cautiously Positive persona after pilot exposure, while also showing that movement occurred in both upward and downward directions.

To aid interpretation, transition probabilities were calculated as row percentages, indicating the proportion of participants from each pre-pilot persona who remained in or moved to each post-pilot persona.

\begin{table}[H]
\centering
\small
\caption{Transition matrix of participants across acceptance personas from pre-pilot to post-pilot.}
\label{tab:cluster_transition}
\renewcommand{\arraystretch}{1.15}
\begin{tabular}{lccccc}
\toprule
\multirow{2}{*}{Acceptance persona} 
& \shortstack{Pre-pilot\\count (\%)} 
& \multicolumn{3}{c}{Post-pilot persona assignment} 
& \shortstack{Post-pilot\\count (\%)} \\
\cmidrule(lr){3-5}
& & Skeptics & Cautiously Positive & Champions & \\
\midrule
Skeptics 
& 35 (28\%) 
& 21 (60\%) 
& 14 (40\%) 
& 0 (0\%) 
& 31 (25\%) \\

Cautiously Positive 
& 70 (56\%) 
& 8 (11\%) 
& 53 (76\%) 
& 9 (13\%) 
& 78 (63\%) \\

Champions 
& 19 (15\%) 
& 2 (10\%) 
& 11 (58\%) 
& 6 (32\%) 
& 15 (12\%) \\
\midrule
Total 
& 124 (100\%) 
& 31 (25\%) 
& 78 (63\%) 
& 15 (12\%) 
& 124 (100\%) \\
\bottomrule
\multicolumn{6}{p{0.95\linewidth}}{\footnotesize \textit{Note.} 
Pre-pilot and post-pilot percentages are calculated relative to the full analytic sample ($N=124$). Percentages in the three transition columns are row percentages and indicate the proportion of participants from each pre-pilot persona who were assigned to each post-pilot persona.} \\
\end{tabular}
\end{table}

\begin{figure}[htbp]
\centering
\includegraphics[width=0.9\textwidth]{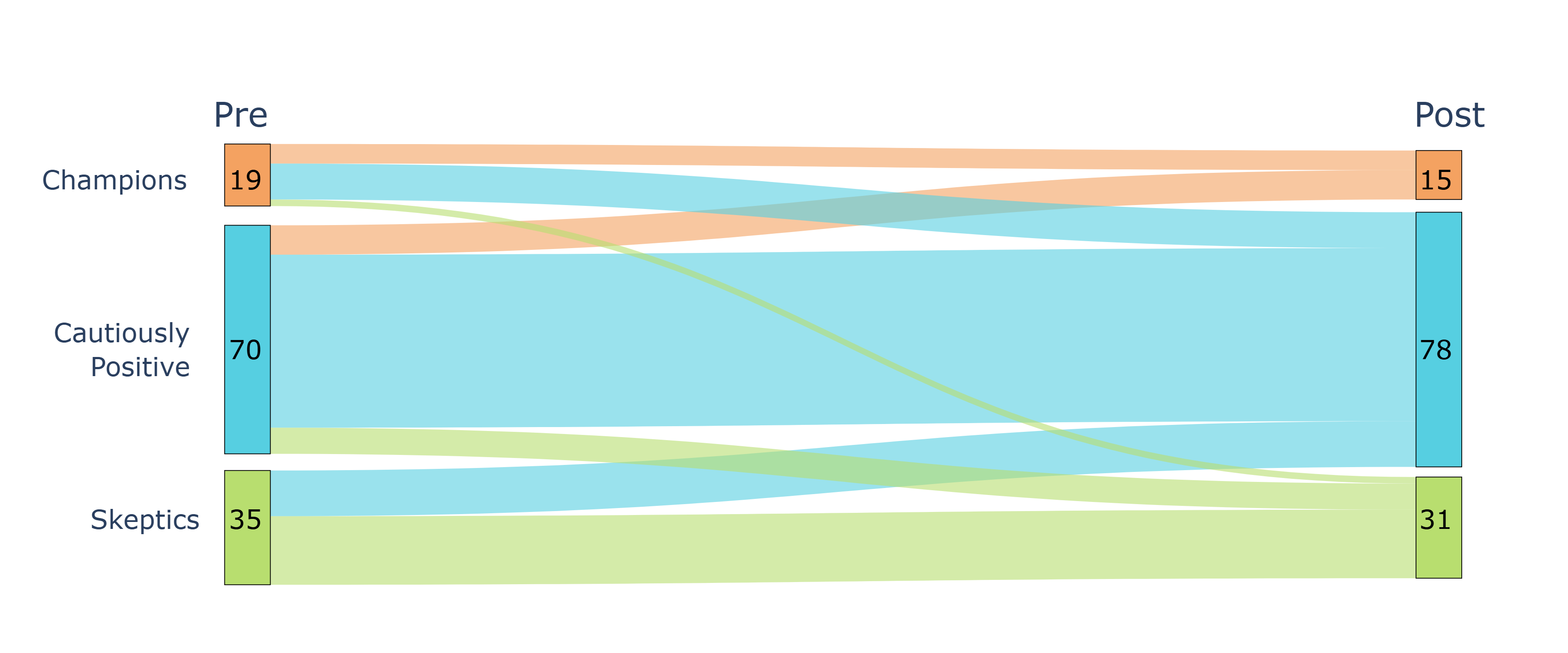}
\caption{Transitions between acceptance personas from pre-pilot to post-pilot 
%based on KMeans clustering. C2: Champions; C1: Cautious Skeptics; C0: Skeptics.
}
\label{fig:Kmeans3_transition}
\end{figure}

Table~\ref{tab:migration_metrics} and Figure~\ref{fig:Kmeans3_transition} further characterize these movements by showing construct-level changes along each migration path. Participants moving upward from Skeptics to Cautiously Positive showed increases across all four constructs, particularly behavioral intention and trust. Participants moving from Cautiously Positive to Champions also showed increases in usefulness, ease of use, behavioral intention, and trust. In contrast, downward movement from Champions to Cautiously Positive or Skeptics was associated with declines in perceived usefulness and trust. Because some paths involved small cell counts, particularly Champions to Skeptics, path-specific estimates should be interpreted descriptively.

\begin{table}[H]
\centering
\caption{Mean pre-pilot, post-pilot, and delta scores in acceptance constructs (PU, PEOU, BI, and TR) for migrating participants along each cluster-migration path
}
\footnotesize
\renewcommand{\arraystretch}{1.35}
\begin{tabular}{c c ccc ccc ccc ccc}
\toprule
\textbf{Migration Path} & \multirow{2}{*}{\textbf{n}}  
& \multicolumn{3}{c}{\textbf{PU}} 
& \multicolumn{3}{c}{\textbf{PEOU}} 
& \multicolumn{3}{c}{\textbf{BI}} 
& \multicolumn{3}{c}{\textbf{TR}} \\
\cmidrule(lr){3-5} \cmidrule(lr){6-8} \cmidrule(lr){9-11} \cmidrule(lr){12-14}
\multicolumn{1}{c}{\textbf{(Pre $\rightarrow$ Post)}} & & \text{Pre} & \text{Post} & $\Delta$ 
  & \text{Pre} & \text{Post} & $\Delta$
  & \text{Pre} & \text{Post} & $\Delta$
  & \text{Pre} & \text{Post} & $\Delta$ \\
\midrule
C0 $\rightarrow$ C0 & 21 & 2.92 & 2.64 & -0.27 & 3.23 & 3.38 & 0.15 & 3.18 & 2.99 & -0.19 & 2.42 & 2.39 & -0.02 \\
C0 $\rightarrow$ C1 & 14 & 3.25 & 3.66 & 0.41 & 3.11 & 3.77 & 0.66 & 2.96 & 3.91 & 0.95 & 2.91 & 3.52 & 0.61 \\
C1 $\rightarrow$ C0 & 8 & 4.22 & 2.97 & -1.25 & 3.41 & 3.09 & -0.31 & 4.00 & 3.34 & -0.66 & 3.25 & 2.91 & -0.34 \\
C1 $\rightarrow$ C1 & 53 & 3.98 & 3.84 & -0.14 & 3.53 & 3.67 & 0.14 & 3.90 & 4.05 & 0.15 & 3.23 & 3.35 & 0.12 \\
C1 $\rightarrow$ C2 & 9 & 3.97 & 4.42 & 0.44 & 3.67 & 4.00 & 0.33 & 3.92 & 4.81 & 0.89 & 3.17 & 3.89 & 0.72 \\
C2 $\rightarrow$ C0 & 2 & 4.75 & 3.25 & -1.50 & 5.00 & 3.75 & -1.25 & 4.62 & 3.50 & -1.12 & 4.12 & 2.25 & -1.88 \\
C2 $\rightarrow$ C1 & 11 & 4.82 & 3.93 & -0.89 & 4.61 & 3.86 & -0.75 & 4.61 & 4.09 & -0.52 & 4.11 & 3.48 & -0.64 \\
C2 $\rightarrow$ C2 & 6 & 4.62 & 4.21 & -0.42 & 4.46 & 4.75 & 0.29 & 4.75 & 4.54 & -0.21 & 4.46 & 4.04 & -0.42 \\
\bottomrule
\end{tabular}
\label{tab:migration_metrics}
\begin{tablenotes}
\scriptsize
\item \textit{Note}. Migration paths with small cell counts, particularly C2$\rightarrow$C0, should not be overinterpreted.
\end{tablenotes}
\end{table}

\subsection{Task-Use Patterns Across Transition Groups}

Table~\ref{tab:copilot_tasks_combined} summarizes pre-pilot anticipated use and post-pilot reported use of Microsoft 365 Copilot across six task categories, stratified by transition group. Because these were multi-select survey items, values represent the proportion of participants in each group who selected each task category. The adjusted $p$-values report within-group pre--post comparisons based on McNemar's test.

Across both transition groups, participants reported high anticipated use prior to the pilot, particularly for communication-related tasks and summarization. After the pilot, task-use patterns differed descriptively between the stable/improved and declined groups. In the stable/improved group, communication-related use remained nearly unchanged, and summarization increased slightly. In contrast, the declined group showed reductions across most task categories, including communication and summarization, suggesting a larger gap between anticipated and reported use among participants whose acceptance profile shifted downward. 

The largest reductions were observed for more structured or output-sensitive tasks. Data/chart-related use declined significantly in both groups, decreasing from 0.71 to 0.44 in the stable/improved group ($\Delta=-0.27$, $p_{\mathrm{adj}}=0.001$) and from 0.86 to 0.38 in the declined group ($\Delta=-0.48$, $p_{\mathrm{adj}}=0.012$). Presentation-related use also declined significantly in the stable/improved group ($\Delta=-0.28$, $p_{\mathrm{adj}}<0.001$), while the corresponding decline in the declined group was not statistically significant after adjustment. Other task categories, including information retrieval, brainstorming, communication, and summarization, did not show statistically significant within-group changes.

Overall, these patterns suggest that participants initially viewed Copilot as broadly applicable across a range of work activities, but post-pilot use coincided with more selective reported task use. Communication and summarization remained relatively stable, particularly among stable/improved participants, whereas data/chart generation and presentation tasks showed larger reductions, indicating that more specialized or accuracy-sensitive tasks may require additional tool refinement, training, or workflow-specific guidance.

\begin{table}[H]
\centering
\scriptsize
\caption{Pre--post changes in Microsoft 365 Copilot task use by transition group.}
\label{tab:copilot_tasks_combined}
\renewcommand{\arraystretch}{1.15}

\resizebox{\textwidth}{!}{%
\begin{tabular}{lcccccccc}
\toprule
\multirow{2}{*}{Task category} 
& \multicolumn{4}{c}{Stable/Improved} 
& \multicolumn{4}{c}{Declined} \\
\cmidrule(lr){2-5} \cmidrule(lr){6-9}
& Pre & Post & $\Delta$ & $p_{\mathrm{adj}}$ 
& Pre & Post & $\Delta$ & $p_{\mathrm{adj}}$ \\
\midrule
Data/charts 
& 0.71 & 0.44 & -0.27 & 0.001* 
& 0.86 & 0.38 & -0.48 & 0.012* \\

Information retrieval 
& 0.68 & 0.65 & -0.03 & 0.791 
& 0.76 & 0.62 & -0.14 & 0.762 \\

Brainstorming 
& 0.55 & 0.46 & -0.10 & 0.371 
& 0.62 & 0.52 & -0.10 & 0.791 \\

Communication 
& 0.83 & 0.84 &  0.02 & 0.791 
& 0.86 & 0.71 & -0.14 & 0.432 \\

Presentations 
& 0.62 & 0.34 & -0.28 & $<0.001$* 
& 0.62 & 0.33 & -0.29 & 0.316 \\

Summarization 
& 0.86 & 0.93 &  0.07 & 0.371 
& 1.00 & 0.86 & -0.14 & 0.432 \\
\bottomrule
\end{tabular}%
}

\vspace{0.3em}

\footnotesize
\raggedright
\textit{Note.} Values are selection rates, interpreted as the proportion of participants in each transition group who selected each task category. 
%Pre-pilot values represent anticipated Copilot use, while post-pilot values represent actual use reported after the pilot. $\Delta$ represents post-pilot minus pre-pilot selection rate. Within-group pre--post comparisons were assessed using McNemar's test. 
Adjusted $p$-values were calculated using the Benjamini--Hochberg false-discovery-rate correction. $^{*}p<0.05$.
%, $^{**}p<0.01$, $^{***}p<0.001$.

\end{table}

\subsection{Concern Patterns Across Transition Groups}
Table~\ref{tab:concerns_combined} summarizes pre--post changes in user concerns regarding Microsoft 365 Copilot across the stable/improved and declined transition groups, together with the corresponding
 within-group statistical comparisons. 

Overall, concerns related to Accuracy and Privacy/data decreased in both groups following the pilot. Accuracy concerns declined from 0.76 to 0.46 in the stable/improved group and from 0.67 to 0.33 in the declined group. The decrease was statistically significant in the stable/improved group ($p_{\mathrm{adj}}<0.001$), while the declined group showed a similar downward pattern that did not remain statistically significant after adjustment ($p_{\mathrm{adj}}=0.075$). Privacy/data concerns also decreased in both groups and were statistically significant for both the stable/improved group ($p_{\mathrm{adj}}<0.001$) and the declined group ($p_{\mathrm{adj}}=0.029$). These results suggest that some initial concerns related to output reliability and data handling became less prominent after hands-on exposure to Copilot.

In contrast, job/skills concerns increased in both transition groups. In the stable/improved group, the proportion of participants selecting this concern increased from 0.20 to 0.43 ($p_{\mathrm{adj}}<0.001$). In the declined group, the increase was larger, rising from 0.05 to 0.48 ($p_{\mathrm{adj}}=0.006$). This pattern suggests that, while some system-level concerns decreased after the pilot, participants became more attentive to potential implications for skill development, over-reliance, critical thinking, or changes in work practices. The larger increase among declined participants indicates that users with less favorable perception shifts may have been especially sensitive to these workforce-related concerns.

The proportion of participants reporting no concerns also increased after the pilot, from 0.16 to 0.38 in the stable/improved group and from 0.29 to 0.43 in the declined group. This increase was statistically significant only in the stable/improved group ($p_{\mathrm{adj}}<0.001$). Taken together, the concern patterns indicate that post-pilot concern patterns were not uniformly higher or lower. Rather, participants appeared to differentiate more clearly between types of risk: concerns about accuracy and privacy/data handling decreased, while concerns about job skills and work practices increased.

These findings complement the task-use patterns reported in Table~\ref{tab:copilot_tasks_combined}. Reductions in some task-use categories should therefore not be interpreted only as reduced interest in Copilot. They may also reflect a more calibrated understanding of where Copilot fits into existing workflows, which tasks require additional oversight, and where further training, governance, or workflow-specific guidance may be needed.

\begin{table}[H]
\centering
\scriptsize
\caption{Pre--post changes in user concerns regarding Microsoft 365 Copilot by transition group.}
\label{tab:concerns_combined}
\renewcommand{\arraystretch}{1.15}

\resizebox{\textwidth}{!}{%
\begin{tabular}{lcccccccc}
\toprule
\multirow{2}{*}{Concern category} 
& \multicolumn{4}{c}{Stable/Improved} 
& \multicolumn{4}{c}{Declined} \\
\cmidrule(lr){2-5} \cmidrule(lr){6-9}
& Pre & Post & $\Delta$ & $p_{\mathrm{adj}}$ 
& Pre & Post & $\Delta$ & $p_{\mathrm{adj}}$ \\
\midrule
Accuracy 
& 0.76 & 0.46 & -0.30 & $<0.001$* 
& 0.67 & 0.33 & -0.33 & 0.075 \\

Privacy/data 
& 0.56 & 0.23 & -0.33 & $<0.001$* 
& 0.57 & 0.19 & -0.38 & 0.029* \\

Job/skills 
& 0.20 & 0.43 &  0.22 & $<0.001$* 
& 0.05 & 0.48 &  0.43 & 0.006* \\

No concerns 
& 0.16 & 0.38 &  0.22 & $<0.001$* 
& 0.29 & 0.43 &  0.14 & 0.375 \\
\bottomrule
\end{tabular}%
}

\vspace{0.3em}

\footnotesize
\raggedright
\textit{Note.} Values are selection rates, interpreted as the proportion of participants in each transition group who selected each concern category.  Concern categories were coded as binary indicators. Within-group pre--post changes were assessed using McNemar's test. Adjusted $p$-values were calculated using the Benjamini--Hochberg false-discovery-rate correction.  $^{*}p<0.05$.% $^{**}p<0.01$, $^{***}p<0.001$.

\end{table}

\subsection{Exploratory Keyword-Based Content Analysis}

To complement the quantitative findings, open-ended survey responses were analyzed using an exploratory keyword-based content analysis. The purpose of this analysis was to examine how frequently participants used language associated with the four TAM-related constructs: perceived usefulness (PU), perceived ease of use (PEOU), behavioral intention (BI), and trust (TR). Because this approach relies on keyword presence and frequency, the results should be interpreted as indicators of construct salience rather than direct measures of sentiment or perception change.

Table~\ref{tab:tam_qualitative_transition} summarizes construct-related keyword presence and mention frequency before and after the pilot, stratified by transition group. PU-related language became substantially more salient in post-pilot responses for both transition groups. In the stable/improved group, the proportion of participants using at least one PU-related keyword increased from 0.22 to 0.83, while the average number of PU-related mentions increased from 0.24 to 1.30 per participant. A similar pattern was observed in the declined group, where PU-related presence increased from 0.29 to 0.71. These patterns suggest that, after hands-on use, participants discussed concrete usefulness, productivity, and task-support applications more frequently. This should not be interpreted as a direct increase in perceived usefulness, since the quantitative survey results showed a decline in aggregate PU scores; rather, the qualitative responses indicate that participants had more specific experiences to reference when discussing usefulness.

PEOU-related language also became more common after the pilot, although it remained less prominent than PU-related language. This suggests that usability and learnability were mentioned by some participants after exposure, but they were not the dominant themes in the open-ended responses. BI-related language showed smaller increases, indicating that some participants referred to future use, continued use, or recommendations, but behavioral intention was less frequently discussed in open-ended comments than usefulness or trust-related issues.

Trust-related keyword presence and mentions declined from pre-pilot to post-pilot in both transition groups. In the stable/improved group, trust-related presence decreased from 0.89 to 0.48, and average trust-related mentions decreased from 3.39 to 0.99. In the declined group, trust-related presence decreased from 0.71 to 0.43, and mentions decreased from 2.76 to 0.90. Because keyword matching does not capture whether a term was used positively or negatively, these results should be interpreted as a change in the salience of trust-related language rather than a direct decline in trust. One possible interpretation is that many pre-pilot comments emphasized anticipated concerns about accuracy, privacy, reliability, and risk, while post-pilot comments shifted toward concrete use cases, workflow fit, and tool limitations.

\begin{table}[H]
\centering
\scriptsize
\caption{Keyword-based content mapping of open-ended responses to TAM-related constructs by transition group.}
\label{tab:tam_qualitative_transition}
\renewcommand{\arraystretch}{1.15}
\resizebox{\textwidth}{!}{%
\begin{tabular}{lcccccccccccc}
\toprule
\multirow{3}{*}{Construct} 
& \multicolumn{6}{c}{Stable/Improved} 
& \multicolumn{6}{c}{Declined} \\
\cmidrule(lr){2-7} \cmidrule(lr){8-13}
& \multicolumn{3}{c}{Presence} 
& \multicolumn{3}{c}{Mentions}
& \multicolumn{3}{c}{Presence} 
& \multicolumn{3}{c}{Mentions} \\
\cmidrule(lr){2-4} \cmidrule(lr){5-7}
\cmidrule(lr){8-10} \cmidrule(lr){11-13}
& Pre & Post & $\Delta$ 
& Pre & Post & $\Delta$ 
& Pre & Post & $\Delta$ 
& Pre & Post & $\Delta$ \\
\midrule
PU   & 0.22 & 0.83 &  0.60 & 0.24 & 1.30 &  1.06 & 0.29 & 0.71 &  0.43 & 0.52 & 1.29 &  0.76 \\
PEOU & 0.03 & 0.22 &  0.19 & 0.04 & 0.48 &  0.44 & 0.05 & 0.29 &  0.24 & 0.05 & 0.67 &  0.62 \\
TR   & 0.89 & 0.48 & -0.42 & 3.39 & 0.99 & -2.40 & 0.71 & 0.43 & -0.29 & 2.76 & 0.90 & -1.86 \\
BI   & 0.06 & 0.14 &  0.08 & 0.08 & 0.18 &  0.11 & 0.00 & 0.14 &  0.14 & 0.00 & 0.24 &  0.24 \\
\bottomrule
\end{tabular}%
}

\vspace{0.3em}
\footnotesize
\raggedright
\textit{Note.} Presence represents the proportion of participants whose open-ended responses contained at least one keyword associated with the construct. Mentions represent the average number of construct-related keyword occurrences per participant. $\Delta$ indicates the post-pilot value minus the pre-pilot value. Keyword-based measures capture the salience of construct-related language and do not indicate whether the statement expressed positive, negative, or neutral sentiment. 
\end{table}

The open-ended responses provide additional context for the quantitative transition results. Four themes were especially salient.

First, participants emphasized workflow-specific usefulness. Several respondents described potential value in drafting, summarization, internal knowledge retrieval, and technical documentation. For example, one technical respondent noted that AI-generated technical documents could provide ``a huge time savings,'' while another described potential value in using Copilot to access institutional knowledge across SharePoint, OneDrive, Teams, and Outlook. These comments help explain why usefulness-related language became more common after the pilot: participants were able to connect the tool to concrete work activities rather than discussing it only in general terms.

Second, participants identified limitations in technical and accuracy-sensitive workflows. Several comments noted challenges with Excel, technical document formatting, recurring errors, or instability in some Microsoft Office applications. One respondent stated that ``the functionality in Excel is lacking,'' while another noted ``recurring errors and instability'' in parts of the Microsoft Office suite. These comments align with the task-use results showing reduced post-pilot selection of data/chart and presentation-related use cases. They suggest that Copilot's perceived value may be strongest for communication and summarization tasks, while more structured or accuracy-sensitive tasks may require additional validation, training, or tool refinement.

Third, participants emphasized the need for role-specific training and support. Several respondents noted that training should be tailored to different job functions rather than delivered uniformly across technical and administrative roles. One respondent explained that ``Engineers'' and ``Administrative functions'' should not always be grouped together because ``the functions are too different to cover together.'' Another respondent suggested a centralized support location where users could post issues and receive answers from both a professional support team and other users. These comments support the broader finding that adoption support should be persona- and workflow-specific rather than one-size-fits-all.

Fourth, respondents expressed interest in customized or expanded AI capabilities. Some participants suggested domain-specific tools for procurement rules, permit review, construction status dashboards, GIS/mapping assistance, policy lookup, or improved search of contract provisions. Others noted that Copilot may need to be supplemented by additional tools for specialized tasks such as image generation, video creation, or template-based workflows. These comments suggest that Microsoft 365 Copilot may be viewed as useful for general productivity tasks, but some users may still need more specialized AI solutions tailored to business-unit workflows.

Overall, the qualitative results reinforce the quantitative pattern of expectation recalibration. Post-pilot comments became more grounded in specific use cases, limitations, and support needs. Participants did not simply become more or less favorable toward Copilot; rather, their comments suggest a more differentiated understanding of where the tool fits well, where it requires human oversight, and where additional training or customization may be needed.

\section{Discussion}
 %, suggesting they anticipate a learning curve. In short, Skeptics doubt the tool's value and accuracy and will likely adopt only if early experiences prove safe and beneficial.
\subsection{Persona Migration and Expectation Recalibration}
At first glance, the aggregate number of participants in each acceptance persona changed only modestly from pre-pilot to post-pilot (Table~\ref{tab:cluster_transition}). Relying only on these aggregate counts could suggest limited movement following the pilot. The transition matrix, however, shows a more dynamic pattern. Among baseline Skeptics, 40\% moved to the Cautiously Positive persona. Among baseline Cautiously Positive users, 76\% remained in the same persona, while 13\% moved upward to Champions and 11\% moved downward to Skeptics. Among baseline Champions, only 32\% remained Champions, while 58\% moved to Cautiously Positive and 10\% moved to Skeptics. Thus, the overall distribution of persona counts masks substantial individual-level movement, which is important for designing follow-up training, support, and governance strategies.

The direction and magnitude of the PU decline are consistent with an expectation-disconfirmation interpretation. Prior to the pilot, participants had limited hands-on experience with Microsoft 365 Copilot: 81\% reported minimal or no prior use (Figure \ref{fig:Likert_Pre}). Pre-pilot PU scores therefore likely reflected optimistic, technology-general expectations rather than task-specific performance assessments. After eight weeks of use within constrained, policy-governed workflows, those expectations were updated against actual experience, a process described in ECM-IT as disconfirmation \cite{bhattacherjee2001understanding}. The significant aggregate decline in PU and the downward migration of the majority of baseline Champions are both consistent with negative disconfirmation: initial expectations exceeded perceived performance. Conversely, upward migration among Skeptics aligns with positive disconfirmation that actual use exceeded the limited expectations this group held at baseline. This expectation-recalibration framing carries practical implications: measured post-pilot PU may reflect a more accurate baseline than pre-pilot PU for forecasting long-term use, rather than signaling a failure of the tool or the pilot intervention.

Table~\ref{tab:migration_metrics} provides additional insight into the construct-level changes underlying these transitions. Upward movement was generally associated with gains in perceived usefulness, trust, and behavioral intention. For example, participants moving from Skeptics to Cautiously Positive ($C0 \rightarrow C1$; $n=14$) showed increases in perceived usefulness (PU: 3.25 to 3.66; $\Delta=+0.41$), trust (TR: 2.91 to 3.52; $\Delta=+0.61$), and behavioral intention (BI: 2.96 to 3.91; $\Delta=+0.95$). Similarly, participants moving from Cautiously Positive to Champions ($C1 \rightarrow C2$; $n=9$) increased across all four constructs, including PU ($\Delta=+0.44$), PEOU ($\Delta=+0.33$), BI ($\Delta=+0.89$), and TR ($\Delta=+0.72$). These patterns suggest that upward movement was associated not only with recognizing the tool's usefulness, but also with increased confidence and intention to use it.

Downward movement followed a different pattern. Participants moving from Champions to Cautiously Positive ($C2 \rightarrow C1$; $n=11$) showed declines in PU ($\Delta=-0.89$), perceived ease of use (PEOU; $\Delta=-0.75$), BI ($\Delta=-0.52$), and TR ($\Delta=-0.64$). The small group moving from Champions to Skeptics ($C2 \rightarrow C0$; $n=2$) showed the largest decline in trust ($\Delta=-1.88$), although this path should be interpreted cautiously because of its small cell count. These path-specific patterns extend the disconfirmation interpretation outlined above to the individual level: upward migration is consistent with positive disconfirmation among users whose limited baseline expectations were exceeded, and downward migration is consistent with negative disconfirmation among users whose high baseline expectations were not met after encountering tool limitations.

Apparent stability also involved smaller but meaningful changes. Participants who remained Cautiously Positive ($C1 \rightarrow C1$; $n=53$) showed a slight decline in PU ($\Delta=-0.14$), but modest gains in PEOU ($\Delta=+0.14$), BI ($\Delta=+0.15$), and TR ($\Delta=+0.12$). Participants who remained Champions ($C2 \rightarrow C2$; $n=6$) showed declines in PU ($\Delta=-0.42$), BI ($\Delta=-0.21$), and TR ($\Delta=-0.42$), despite a small increase in PEOU ($\Delta=+0.29$). These patterns suggest that remaining in the same persona does not imply unchanged perceptions. Even stable persona membership may involve recalibration of expectations after hands-on use.

Across the migration paths, trust appears to be a key dynamic construct. Trust gains accompanied upward movement, while trust losses were prominent among downward movers, particularly among Champions who shifted to less favorable personas. The findings suggest that sustained adoption depends not only on perceived usefulness and ease of use, but also on maintaining appropriately calibrated trust through clear use cases, verification routines, reliability guardrails, and post-training support.

\subsection{Task-Use Calibration and Workflow Fit}
The task-use results in Table~\ref{tab:copilot_tasks_combined} suggest that hands-on experience helped participants refine their expectations about where Microsoft 365 Copilot was most useful. Prior to the pilot, participants in both transition groups anticipated using Copilot across a broad range of tasks, including communication, summarization, data/chart generation, information retrieval, brainstorming, and presentations. After the pilot, however, reported use became more selective. Communication and summarization remained comparatively stable, particularly among participants in the stable/improved transition group, whereas data/chart generation declined significantly in both transition groups and presentation-related use declined significantly among stable/improved participants.
These patterns suggest that participants may have initially overestimated the tool's usefulness for more structured or output-sensitive tasks. Data/chart generation and presentation development often require higher levels of accuracy, formatting control, domain knowledge, or verification than routine drafting and summarization tasks. The larger descriptive reductions observed among declined participants may reflect a greater mismatch between initial expectations and actual workflow fit, especially for users whose work required more technical precision or specialized outputs.

Participant role and work context may help explain these task-use patterns. Technical users may be more likely to test Copilot on analytical, engineering, spreadsheet, or domain-specific tasks where output accuracy and verification requirements are higher. Administrative and support users, by contrast, may experience more immediate value from communication-oriented functions such as drafting, summarizing, and organizing information within Microsoft 365 applications. Thus, the observed reduction in some task categories should not be interpreted simply as reduced interest. Rather, it suggests a more calibrated understanding of which workflows are well supported by the current tool and which may require additional training, validation procedures, or more specialized AI capabilities.

\subsection{Concern Recalibration After Hands-On Use}
The concern patterns in Table~\ref{tab:concerns_combined} suggest that hands-on use helped participants differentiate between system-level risks and workforce-related implications. Accuracy and privacy/data concerns decreased after the pilot, while job/skills concerns increased in both transition groups. This pattern indicates that participants did not simply become less concerned overall. Instead, their concerns shifted from whether the system could be used safely and reliably toward how AI assistance might affect skills, judgment, and work practices.

Role and work context may help explain these patterns. Technical users may be more likely to test Copilot in accuracy-sensitive or domain-specific workflows where verification and professional judgment are critical. Administrative and support users may experience more immediate value in communication, summarization, and routine productivity tasks. Together, these findings suggest that governance and training should address more than privacy and accuracy. While technical guardrails and human review procedures remain important, agencies should also address workforce-facing concerns by clarifying appropriate use cases, reinforcing the role of professional judgment, and helping employees understand how AI assistance should complement rather than replace domain expertise. 
The rise in job/skills concerns warrants attention beyond routine training design. Research on AI-induced skill anxiety in knowledge work suggests that employees who perceive AI as capable of performing core elements of their role (e.g., drafting, summarizing, information retrieval) may experience threats to professional identity and role clarity, particularly when the boundary between AI assistance and AI substitution is ambiguous~\cite{raisch2021artificial}. In the present study, the increase was largest among declined participants (43 percentage points), consistent with the interpretation that users who encountered limitations or unmet expectations may have also confronted more acute uncertainty about how to position Copilot within their professional practice. This finding aligns with recent research showing that workforce concerns about skill displacement tend to increase, not decrease, with greater AI exposure, at least in early adoption periods~\cite{dellacqua2023navigating}.
Agencies should therefore treat rising job/skills concerns not as evidence of resistance to be overcome, but as a signal that employees are actively working through questions of role fit, professional judgment, and appropriate reliance, questions that training and governance can address but not eliminate through reassurance alone.

\subsection{Implications for DOT rollout and governance}
The migration results suggest that a uniform training and rollout strategy is unlikely to support all users equally. Employees entered the pilot with different levels of perceived usefulness, ease of use, behavioral intention, and trust, and their post-pilot trajectories also differed. Some initially skeptical users moved toward more favorable perceptions, while some initially enthusiastic users revised their expectations downward after hands-on use. These patterns suggest that generative AI adoption in a DOT should be managed as an adaptive process rather than a one-time deployment. 

A more effective approach is to link interventions to training dosage, peer social learning, and error governance while monitoring success with construct‑level thresholds. For employees who begin as Skeptics, the primary implementation need is trust formation. Agencies can support this group by presenting a limited set of verified, low-risk, domain-specific use cases and by demonstrating how AI-generated outputs should be checked before use. Examples should show not only successful outputs, but also common failure modes, such as inaccurate summaries, incomplete responses, or unsupported claims. This can help users develop calibrated trust: sufficient confidence to experiment with the tool, but not so much confidence that human review is bypassed.
For Cautiously Positive users, the main need is workflow fluency. This group already shows moderate acceptance but may require practical support to translate general interest into routine use. Short micro-learning modules, prompt-writing practice, office hours, and peer examples may help these users identify where Copilot fits into daily tasks such as drafting, summarizing, meeting preparation, and information retrieval. Rather than providing generic training, DOTs should tailor examples to specific functions, such as engineering documentation, procurement, planning, communications, or administrative support.
Champions require a different form of support. Although these users begin with high expectations, the migration results show that enthusiasm can decline after hands-on experience if the tool does not meet expectations or if errors require substantial correction. Agencies can use Champions as peer mentors or prompt-library contributors, but they should also reinforce verification routines and responsible-use expectations. In this way, Champions can help diffuse practical knowledge while also modeling appropriate human oversight.

The task-use and concern patterns further suggest that AI governance should address both technical and workforce-facing risks. Participants appeared to recalibrate their expectations after use: communication and summarization tasks remained comparatively stable, while more structured or accuracy-sensitive tasks, such as data/chart generation and presentation development, showed larger reductions. At the same time, concerns about accuracy and privacy/data handling decreased, while job/skills concerns increased. These findings indicate that governance should go beyond access control and privacy safeguards. Agencies should also provide guidance on when Copilot is appropriate, when outputs require additional verification, and how AI use should complement rather than replace professional judgment. Construct-level measures such as perceived usefulness, perceived ease of use, behavioral intention, and trust can serve as diagnostic indicators for monitoring rollout progress. Rather than treating acceptance surveys as a one-time evaluation, DOTs can use periodic measurement to identify groups that are gaining confidence, groups that are plateauing, and groups whose expectations are declining. This approach can help agencies target follow-up training, revise use-case guidance, and adjust governance practices as employees gain experience.

Finally, although OMB Memorandum M-24-10 applies to federal agencies, it provides a useful governance benchmark for public-sector AI deployment, particularly around safeguards, accountability, and risk management \cite{OMB2024AIGovernance}. For state DOTs, the broader lesson is that durable adoption depends not only on demonstrating usefulness, but also on continuously cultivating appropriately calibrated trust through training, human review expectations, risk communication, and workflow-specific support.

\subsection{Limitations and Future Directions}
This study captures the early stages of generative AI adoption during an eight-week pilot. Although the two-wave design allowed us to examine how perceptions changed after training and hands-on use, it does not show whether observed gains or losses in perceived usefulness, ease of use, behavioral intention, or trust would persist over a longer implementation period.
The study also reflects the context of a single state DOT. Organizational culture, governance practices, training approach, and workforce composition may differ from those of other transportation agencies. Therefore, the exact persona proportions and migration rates should not be assumed to generalize directly to other DOTs or public agencies. 
All four acceptance constructs were measured using self-reported Likert-scale responses. These measures are appropriate, but they may not fully reflect actual use behavior. Self-reports may also be affected by social desirability, or response styles. 
Migration paths with very small counts, particularly the transition from Champions to Skeptics, should not be overinterpreted as stable subgroup patterns. In addition, the exploratory qualitative analysis relied on keyword matching, so nuanced semantic meaning and sentiment may not have been fully captured.
%Finally, we anchored the post-survey data to the centroids from the pre-pilot survey K-means model. This simplifies trajectory analysis but assumes that the underlying persona structure itself did not change. 
Future work should extend the study design in several directions. A third survey wave at six or twelve months would help determine whether expectation recalibration, trust shifts, and persona migration persist after the initial pilot period. Replication across multiple DOTs would test whether the three-persona structure generalizes across different public-sector settings. Finally, linking survey responses to objective telemetry, such as frequency of Copilot use, application-specific use, prompt volume, task type, and correction or verification behavior, would help determine whether behavioral intention translates into sustained and effective use.

\section{Conclusion}
This study provides one of the first longitudinal examinations of how employees in a transportation agency responded to LLM assistants in routine office work during an eight-week pilot. By combining a matched two-wave Technology Acceptance Model (TAM) survey with persona-based transition analysis, the study moved beyond aggregate pre--post averages and identified three baseline acceptance personas: Skeptics, Cautiously Positive users, and Champions.
At the aggregate level, perceived usefulness declined significantly after hands-on use, suggesting that participants recalibrated their initial expectations about Copilot's value for their work. In contrast, perceived ease of use, behavioral intention, and trust showed only small, non-significant changes. However, the persona transition analysis revealed a more dynamic pattern than aggregate averages alone. Forty percent of baseline Skeptics moved up to the Cautiously Positive persona, while 68\% of baseline Champions moved to a less enthusiastic persona, including 58\% who shifted to Cautiously Positive and 10\% who shifted to Skeptics. Overall, the post-pilot distribution consolidated around the Cautiously Positive persona, which increased from 56\% to 63\% of the analytic sample.
Migration-path results further showed that upward movement was associated with gains in perceived usefulness, behavioral intention, and trust. Conversely, downward movement, particularly among Champions, was associated with declines in perceived usefulness and trust. These patterns suggest that trust is a key dynamic construct in generative AI adoption. Usefulness and intention may support initial interest, while sustained adoption appears to depend on maintaining appropriately calibrated trust through reliable outputs, verification practices, and continued support.
The task-use and concern patterns also indicate that participants developed a more differentiated understanding of the tool after hands-on experience. Communication and summarization remained comparatively stable use cases, while more structured or accuracy-sensitive tasks, such as data/chart generation and presentation development, showed larger reductions. At the same time, concerns about accuracy and privacy/data handling decreased, whereas job/skills concerns increased. These findings suggest that public agencies should not interpret reduced use in some task categories simply as reduced interest; rather, it may reflect a more realistic understanding of where generative AI tools fit within existing workflows and where additional training, governance, or human oversight is needed.
For DOT implementation, the findings point to the value of persona-specific adoption support. Employees who begin with low trust may benefit from verified use cases, demonstrations of human review, and low-risk opportunities to build confidence. Cautiously Positive users may need workflow-specific training, prompt practice, and peer examples to translate general interest into routine use. Champions may serve as mentors or prompt-library contributors, but they also need trust-calibration safeguards to prevent overconfidence and reinforce appropriate verification practices.

%\section{Acknowledgments}
%The authors gratefully acknowledge the North Carolina Department of Transportation (NCDOT) for supporting this study and for providing the organizational access, subject matter expertise, and operational context necessary to conduct the research. The authors especially thank Ryan Brumfield and Caitlyn Mabry of NCDOT's Office of Strategic Initiatives and Program Support for their guidance and assistance throughout the pilot. The authors also thank the Ernst \& Young (EY) team for supporting pilot implementation, training activities, survey instrument development, and survey distribution. Finally, the authors are grateful to the NCDOT employees who participated in the pilot, completed the surveys, and shared their experiences with generative AI tools.
%The views, findings, interpretations, and conclusions expressed in this paper are those of the authors and do not necessarily reflect the official views, policies, or positions of NCDOT, EY, or any other organization involved in the pilot. 

\bibliographystyle{elsarticle-num-names} 
\bibliography{Refs}

%\lipsum[1-10]
%\end{thebibliography}
\end{document}